\documentclass[11pt,a4paper]{article}

\usepackage{bm}
\usepackage[top=15truemm,bottom=20truemm,left=20truemm,right=20truemm]{geometry}
\usepackage[colorlinks=true,urlcolor=blue,anchorcolor=black,citecolor=blue,linkcolor=black,filecolor=black,menucolor=black,linktocpage=true,pdfproducer=medialab,pdfa=true]{hyperref}
\usepackage{graphicx}
\usepackage{cite}
\usepackage{amsmath,latexsym,amssymb,mathrsfs,ascmac,physics,mathtools,slashed,xcolor}
\numberwithin{equation}{section}
\usepackage[affil-it]{authblk}
\newcommand{\email}[1]{\thanks{\href{mailto:#1}{\texttt{#1}}}}

\usepackage{amsthm}
\allowdisplaybreaks[1]

\title{Separation of variables for coupled radiative perturbations of Kerr--Newman--de Sitter black holes}
\author{Daisuke Yoshida
\email{yoshida.daisuke.k9@f.mail.nagoya-u.ac.jp}}
\affil{Institute for Advanced Research, Nagoya University,
Nagoya 464-8602, Japan}
\date{}

\begin{document}

\maketitle

\begin{abstract}
We extend Hintz's separation of variables for radiative
gravitational and electromagnetic perturbations from Kerr--Newman
to Kerr--Newman--de Sitter black holes.
On subextremal backgrounds with nonzero rotation and charge,
an explicit algebraic matrix transformation yields coupled
two-component radial and angular ordinary differential systems
linked by a common separation constant.
The transformation is independent of the
frequency and azimuthal number and applies to both signs of spin
weight. For real frequencies, we establish a complete angular mode
decomposition. We also translate regularity at the future black hole
outer and cosmological horizons into ingoing and outgoing boundary
conditions, respectively, for the separated radial systems.
These results show that exact separation of the coupled radiative
equations extends beyond asymptotically flat Kerr--Newman spacetime.
\end{abstract}

\setcounter{tocdepth}{2}
\tableofcontents

\section{Introduction}
\label{sec:introduction}

Separation of variables reduces black hole perturbation equations
from partial differential equations to radial and angular ordinary
differential equations, providing a basis for studying wave scattering
and quasinormal modes. Carter's separable Einstein--Maxwell geometries
connected the structure of rotating spacetimes with the separation of
the Hamilton--Jacobi and scalar wave equations, including a cosmological
constant~\cite{Carter:1968ks}.
Teukolsky obtained separable equations for gravitational and electromagnetic
perturbations of the Kerr black hole~\cite{Teukolsky:1972my},
while Page separated the charged Dirac equation on the Kerr--Newman
background~\cite{Page:1976jj}. These results illustrate that rotation
and electric charge can coexist with separability, but the answer
depends on the field equations under consideration.

On a charged background, gravitational and electromagnetic
perturbations are dynamically coupled. For Kerr--Newman, retaining
this coupling made a direct extension of the decoupled Kerr equations
a longstanding problem.
Dudley and Finley studied separable wave equations for individual spin
fields on type D backgrounds~\cite{Dudley:1977zz,Dudley:1978vd}.
The associated Dudley--Finley approximation provides separated equations
for Kerr--Newman perturbations but does not describe the full coupled
gravitational and electromagnetic dynamics. Its quasinormal mode
predictions and limitations were investigated by Berti and
Kokkotas~\cite{Berti:2005eb} and more recently compared with the
coupled spectrum by Saha and Silva~\cite{Saha:2025nsg}.

For generic rotation and charge, numerical studies determined
quasinormal mode spectra by solving the coupled Kerr--Newman
equations in both radial and angular variables
simultaneously~\cite{Dias:2015wqa,Dias:2022oqm}.
A tensorial formulation of the coupled system
provided a route to analysis directly in
physical space, without relying on separation~\cite{Giorgi:2020ujd}.
Perturbative treatments also obtained mode spectra while retaining
the gravitational and electromagnetic coupling, using expansions
in small rotation~\cite{Pani:2013ija,Pani:2013wsa}
or small charge at arbitrary rotation~\cite{Mark:2014aja}.
Near extremality, a near-horizon expansion yielded separated
equations and analytic approximations to quasinormal mode
frequencies~\cite{Dias:2022oqm}.

Hintz recently obtained an exact separation
of the radiative first-order systems using chiral variables and a matrix
integrating factor~\cite{Hintz:2026jiw}. The separated equations have
since been used for numerical quasinormal mode calculations~\cite{Destounis:2026ljd}
and analytic calculations of static tidal responses~\cite{Chakraborty:2026kyf};
a charge-regular formulation and the relation to physical radiative
amplitudes were developed in Ref.~\cite{Motohashi:2026wac}.
Here separation means radial and angular ordinary differential equations
for each Fourier mode; the gravitational and electromagnetic variables
may remain coupled within each system.

With a positive cosmological constant, separable equations for
Kerr--de Sitter perturbations were developed by Khanal~\cite{Khanal:1983vb}
and used in the horizon analysis of Chambers and Moss~\cite{Chambers:1994ap}.
Suzuki, Takasugi, and Umetsu expressed the separated equations in Heun
form and constructed analytic solutions for Kerr--de Sitter and
applicable test fields on Kerr--Newman--de
Sitter~\cite{Suzuki:1998vy,Suzuki:1999nn}.
The Heun formulation has been applied to quasinormal modes
and wave scattering~\cite{Hatsuda:2020sbn,Motohashi:2021zyv,Furugori:2026bqu}.
These studies concern separable Teukolsky equations and do not include
the coupled gravitational and electromagnetic radiative system
considered here.
For coupled gravitational and electromagnetic perturbations of
Kerr--Newman--de Sitter, Weller et al.~\cite{Weller:2026bfn}
developed a radiative reconstruction procedure and established
parity isospectrality.

We ask whether the coupled gravitational and electromagnetic
radiative fields can still be decomposed into independent angular
sectors in the presence of a positive cosmological constant.
The horizon boundary conditions must also be compatible with
this decomposition.

We extend Hintz's exact separation from Kerr--Newman
to Kerr--Newman--de Sitter.
The central step is to construct a matrix transformation $\mathsf X$
compatible with both the radial and angular dependence of the coupled
equations.
The nontrivial requirement is to satisfy the transport and coupling
conditions simultaneously with a matrix that remains regular and
invertible throughout
the exterior region including both horizons and the rotation axis.
Building on Hintz's method, we allow a sum of a purely radial
and a purely angular term to remain in the transformed derivative
equations. This condition admits an explicit algebraic transformation
and still yields separated radial and angular systems.
The construction does not require an expansion in the cosmological
constant, rotation, or charge.
The separation applies to the radiative systems for both signs
of spin weight on subextremal backgrounds with nonzero rotation
and charge. We also establish angular completeness for real
frequencies and derive horizon boundary conditions for the
separated radial systems.

Shortly after the first version of this work,
Zhao presented a particularly simple algebraic separation
matrix for Kerr--Newman--(anti-)de Sitter
black holes~\cite{Zhao:2026separation}.
Sec.~\ref{app:zhao} gives this matrix and its associated
coefficients in our conventions, and relates it to our
transformation through the freedom described in
Sec.~\ref{sec:local-kn-limit}.

Sec.~\ref{sec:background} defines the Newman--Penrose quantities and derives
the radiative equations and their common chiral form.
Sec.~\ref{sec:transformation} explains the separation conditions,
states the main result, and constructs the matrix satisfying those
conditions. Secs.~\ref{sec:separated} and \ref{sec:horizons} treat
angular completeness and physical horizon conditions.
Sec.~\ref{sec:discussion} summarizes the results and discusses
extensions and open problems.

We use signature $(-+++)$ and units in which $c=G=k_{\mathrm e}=1$,
where $k_{\mathrm e}$ is Coulomb's constant.
Our curvature convention is
$R^\rho{}_{\sigma\mu\nu}
\coloneqq\partial_\mu\Gamma^\rho_{\nu\sigma}
 -\partial_\nu\Gamma^\rho_{\mu\sigma}
 +\Gamma^\rho_{\mu\lambda}\Gamma^\lambda_{\nu\sigma}
 -\Gamma^\rho_{\nu\lambda}\Gamma^\lambda_{\mu\sigma}$, 
$R_{\mu\nu}\coloneqq R^\rho{}_{\mu\rho\nu}$.
Here $\Gamma^\rho_{\mu\nu}$ denotes the Christoffel symbols.

\section{Radiative equations on Kerr--Newman--de Sitter}
\label{sec:background}
\label{app:np}

We consider linear gravitational and electromagnetic
perturbations of Kerr--Newman--de Sitter spacetime.
The full metric and electromagnetic field satisfy
\begin{equation}
 R_{\mu\nu}-\frac12 Rg_{\mu\nu}
 +\Lambda g_{\mu\nu}
 =2\left(F_{\mu\rho}F_\nu{}^\rho
       -\frac14g_{\mu\nu}F_{\rho\sigma}F^{\rho\sigma}\right),
 \qquad \nabla_\mu F^{\mu\nu}=0,\qquad \bm{d}\bm{F}=0.
 \label{eq:bg-field-equations}
\end{equation}
These equations imply $R=4\Lambda$.
We linearize them about the Kerr--Newman--de Sitter background
with $\Lambda$ held fixed.
We define the Newman--Penrose quantities and evaluate them on this background,
derive the radiative equations for positive and negative spin weights,
and put both systems in a common chiral form.

\subsection{Kerr--Newman--de Sitter background}
\label{sec:geometry}

We use Boyer--Lindquist coordinates $(t,r,\theta,\phi)$, with $\phi$ of period $2\pi$.
In the time and azimuthal normalization of
Refs.~\cite{Suzuki:1998vy,Furugori:2026bqu},
the metric and electromagnetic potential are
\begin{align}
 \bm{g}={}&-\frac{\Delta_r}{\Xi^2\Sigma}(\bm{d}t-a\sin^2\theta\,\bm{d}\phi)^2
 +\frac{\Sigma}{\Delta_r}(\bm{d}r)^2+\frac{\Sigma}{\Delta_\theta}(\bm{d}\theta)^2
 +\frac{\Delta_\theta\sin^2\theta}{\Xi^2\Sigma}
       \bigl(a\,\bm{d}t-(r^2+a^2)\bm{d}\phi\bigr)^2,
 \label{eq:bg-metric}\\
 \bm{A}={}&-\frac{Qr}{\Xi\Sigma}
                       (\bm{d}t-a\sin^2\theta\,\bm{d}\phi),\qquad
 \bm{F}=\bm{d}\bm{A}.
 \label{eq:bg-potential}
\end{align}
Here $M,a,Q$ denote the mass, rotation, and electric-charge parameters appearing in the metric. The de Sitter radius $\ell$ has dimensions of length and satisfies $\Lambda=3/\ell^2$. The functions entering the metric are
\begin{align}
 &\Xi\coloneqq 1+\ell^{-2}a^2,\qquad \Sigma(r,\theta)\coloneqq r^2+a^2\cos^2\theta,
 \label{eq:bg-notation}\\
 &\Delta_r(r)\coloneqq (r^2+a^2)(1-\ell^{-2}r^2)-2Mr+Q^2,\qquad
 \Delta_\theta(\theta)\coloneqq 1+\ell^{-2}a^2\cos^2\theta.
 \label{eq:bg-polynomials}
\end{align}
We assume throughout
\begin{equation}
 M>0,\qquad a>0,\qquad Q\ne0,\qquad \ell>0,
 \label{eq:bg-parameter-assumptions}
\end{equation}
and impose the subextremal condition that $\Delta_r$ has three distinct positive roots $0<r_{\mathrm i}<r_{\mathrm o}<r_{\mathrm c}$, with $\Delta_r>0$ on $(r_{\mathrm o},r_{\mathrm c})$. These are the black hole inner, black hole outer, and cosmological horizon radii, respectively.
Explicit conditions on the parameters for this horizon structure
are summarized in Appendix~B of Ref.~\cite{Furugori:2026bqu}.
The black hole inner and outer horizon radii
$r_{\mathrm i}$ and $r_{\mathrm o}$ are conventionally
denoted by $r_-$ and $r_+$, respectively. The exterior domain in the $(r,\theta)$ coordinates is
given by
\begin{equation}
 r_{\mathrm o}<r<r_{\mathrm c},\qquad 0<\theta<\pi.
 \label{eq:bg-exterior}
\end{equation}
The exterior region including the black hole outer horizon,
the cosmological horizon, and the rotation axis is given by
$r_{\mathrm o}\le r\le r_{\mathrm c}$ and $0\le\theta\le\pi$.

\subsection{Newman--Penrose quantities}
\label{app:np-identities}

\paragraph{Definitions.}
We follow the Newman--Penrose conventions of Ref.~\cite{Stephani:2003}.
The definitions below apply to the full fields and null frame,
before linearization; their Kerr--Newman--de Sitter background values are given afterward.
Let $(\bm{e}_k,\bm{e}_l,\bm{e}_m,\bm{e}_{\bar m})
\coloneqq(\bm{k}_{\mathrm{NP}},\bm{l}_{\mathrm{NP}},\bm{m}_{\mathrm{NP}},
\overline{\bm{m}}_{\mathrm{NP}})$ be a null tetrad normalized by
$\bm{k}_{\mathrm{NP}}\cdot\bm{l}_{\mathrm{NP}}=-1$ and
$\bm{m}_{\mathrm{NP}}\cdot\overline{\bm{m}}_{\mathrm{NP}}=1$.
Tensor indices in this section refer to this frame;
$C_{ijkl}$ denotes the all-lowered Weyl tensor.
We define the Weyl scalars $\Psi_i$ and Maxwell scalars $\Phi_i$ by
\begin{align}
 &\Psi_0\coloneqq C_{k m k m},\quad \Psi_1\coloneqq C_{k l k m},\quad
 \Psi_2\coloneqq C_{k m \bar m l},\quad \Psi_3\coloneqq C_{k l \bar m l},\quad \Psi_4\coloneqq C_{l \bar m l \bar m},
 \label{eq:np-weyl-definitions}\\
 &\Phi_0\coloneqq F_{k m},\qquad \Phi_1\coloneqq (F_{k l}+F_{\bar m m})/2,\qquad \Phi_2\coloneqq F_{\bar m l}.
 \label{eq:np-maxwell-definitions}
\end{align}
The traceless part of the Ricci tensor can be described by the Newman--Penrose Ricci scalars defined by
\begin{equation}
\begin{aligned}
 \Phi_{00}&\coloneqq \tfrac12 R_{k k},&
 \Phi_{01}&\coloneqq \tfrac12 R_{k m},\\
 \Phi_{02}&\coloneqq \tfrac12 R_{m m},&
 \Phi_{11}&\coloneqq \tfrac14(R_{k l}+R_{m\bar m}),\\
 \Phi_{12}&\coloneqq \tfrac12 R_{l m},&
 \Phi_{22}&\coloneqq \tfrac12 R_{l l}.
\end{aligned}
\label{eq:np-ricci-definitions}
\end{equation}
The remaining components follow from
$\Phi_{ji}=\overline{\Phi_{ij}}$.
For the Einstein--Maxwell system with cosmological constant,
Eq.~\eqref{eq:bg-field-equations}, these scalars satisfy
$\Phi_{ij}=2\Phi_i\bar\Phi_j$.
Define the Ricci rotation coefficients by
\begin{equation}
 \Gamma_{ijk}\coloneqq \bm{g}(\bm{e}_i,\nabla_{\bm{e}_k}\bm{e}_j)
             =(e_i)_\nu(e_j)^\nu{}_{;\mu}(e_k)^\mu.
\end{equation}
Our Newman--Penrose spin coefficients are defined by
\begin{equation}
\begin{gathered}
\begin{aligned}
 \kappa_{\mathrm{NP}}&\coloneqq -\Gamma_{m k k},&\sigma_{\mathrm{NP}}&\coloneqq -\Gamma_{m k m},&
 \lambda_{\mathrm{NP}}&\coloneqq \Gamma_{\bar m l \bar m},&\nu_{\mathrm{NP}}&\coloneqq \Gamma_{\bar m l l},\\
 \rho_{\mathrm{NP}}&\coloneqq -\Gamma_{m k \bar m},&\tau_{\mathrm{NP}}&\coloneqq -\Gamma_{m k l},&
 \mu_{\mathrm{NP}}&\coloneqq \Gamma_{\bar m l m},&\pi_{\mathrm{NP}}&\coloneqq \Gamma_{\bar m l k},
\end{aligned}\\
\begin{aligned}
 \epsilon_{\mathrm{NP}}&\coloneqq \tfrac12(\Gamma_{k l k}-\Gamma_{m \bar m k}),&
 \gamma_{\mathrm{NP}}&\coloneqq \tfrac12(\Gamma_{k l l}-\Gamma_{m \bar m l}),\\
 \beta_{\mathrm{NP}}&\coloneqq \tfrac12(\Gamma_{k l m}-\Gamma_{m \bar m m}),&
 \alpha_{\mathrm{NP}}&\coloneqq \tfrac12(\Gamma_{k l \bar m}-\Gamma_{m \bar m \bar m}).
\end{aligned}
\end{gathered}
\end{equation}

The directional derivatives are
\begin{equation}
 D\coloneqq k_{\mathrm{NP}}^\mu\nabla_\mu,\qquad
 \Delta\coloneqq l_{\mathrm{NP}}^\mu\nabla_\mu,\qquad
 \delta\coloneqq m_{\mathrm{NP}}^\mu\nabla_\mu,\qquad
 \bar\delta\coloneqq \bar m_{\mathrm{NP}}^\mu\nabla_\mu.
\end{equation}

The definitions above allow us to use the perturbation
variables of Ref.~\cite{Hintz:2026jiw} with signature $(-+++)$,
although that reference uses $(+---)$.
The Weyl and Maxwell scalars and the Newman--Penrose spin coefficients
retain their numerical values, so the signature change
introduces no additional signs in the radiative equations
or the chiral system.

\paragraph{Kerr--Newman--de Sitter background values.}
For the tetrad and the separation analysis below, we introduce
the auxiliary quantities
\begin{align}
 &z(r,\theta)\coloneqq r+ia\cos\theta,\qquad
 w(r,\theta)\coloneqq r-ia\cos\theta,
 \nonumber\\
 &u(r)\coloneqq \sqrt{\Delta_r},\qquad
 h(\theta)\coloneqq \sqrt{\Delta_\theta},\qquad
 v(\theta)\coloneqq a h(\theta)\sin\theta.
 \label{eq:bg-square-roots}
\end{align}
These definitions give $\Sigma=zw$.
The functions $u,v$ are positive
for $r_{\mathrm o}<r<r_{\mathrm c}$ and
$0<\theta<\pi$, respectively.

On the Kerr--Newman--de Sitter background, we choose the tetrad
\begin{align}
 \bm{k}_{\mathrm{NP}}={}&\frac{\Xi(r^2+a^2)}{\Delta_r}\bm{\partial}_t+\bm{\partial}_r
                  +\frac{\Xi a}{\Delta_r}\bm{\partial}_\phi,
 \label{eq:bg-tetrad-k}\\
 \bm{l}_{\mathrm{NP}}={}&\frac{\Xi(r^2+a^2)\bm{\partial}_t-\Delta_r\bm{\partial}_r+\Xi a\bm{\partial}_\phi}{2\Sigma},
 \label{eq:bg-tetrad-l}\\
 \bm{m}_{\mathrm{NP}}={}&\frac1{\sqrt2z}
 \left(\frac{i\Xi a\sin\theta}{h}\bm{\partial}_t+h\bm{\partial}_\theta
                     +\frac{i\Xi}{h\sin\theta}\bm{\partial}_\phi\right).
 \label{eq:bg-tetrad}
\end{align}

Evaluating the Maxwell, Weyl, and traceless Ricci scalars
in this tetrad gives the nonvanishing background values
\begin{equation}
 \Phi_1=\frac{Q}{2w^2},\qquad
 \Psi_2=\frac{Q^2-Mz}{zw^3},\qquad
 \Phi_{11}=\frac{Q^2}{2\Sigma^2},
 \label{eq:bg-curvature-scalars}
\end{equation}
while all other scalars in these three groups vanish.
The conditions $\Psi_0=\Psi_1=0$ and $\Psi_3=\Psi_4=0$
show that $\bm{k}_{\mathrm{NP}}$ and $\bm{l}_{\mathrm{NP}}$
are aligned with the two repeated principal null directions,
respectively. Thus, the background is of Petrov type D,
and this is a principal null tetrad.
All three scalars in Eq.~\eqref{eq:bg-curvature-scalars}
are independent of $\Lambda$ and have the same expressions
as in Kerr--Newman.

Direct differentiation of Eqs.~\eqref{eq:bg-tetrad-k}--\eqref{eq:bg-tetrad} yields
\begin{equation}
 \begin{gathered}
 \rho_{\mathrm{NP}}=-\frac1w,\qquad
 \mu_{\mathrm{NP}}=-\frac{\Delta_r}{2zw^2},\qquad
 \gamma_{\mathrm{NP}}=\mu_{\mathrm{NP}}+\frac{\Delta_r'}{4zw},
 \\[3pt]
 \beta_{\mathrm{NP}}=\frac{h\cot\theta+\partial_\theta h}{2\sqrt2z},\qquad
 \pi_{\mathrm{NP}}=\frac{iv}{\sqrt2w^2},\qquad
 \tau_{\mathrm{NP}}=-\frac{iv}{\sqrt2zw},\qquad
 \alpha_{\mathrm{NP}}=\pi_{\mathrm{NP}}-\bar\beta_{\mathrm{NP}},
 \\[3pt]
 \kappa_{\mathrm{NP}}=\sigma_{\mathrm{NP}}
 =\lambda_{\mathrm{NP}}=\nu_{\mathrm{NP}}
 =\epsilon_{\mathrm{NP}}=0.
 \end{gathered}
 \label{eq:np-background-coefficients}
\end{equation}
A prime on $\Delta_r$ denotes $d/dr$.

\subsection{Radiative equations}
\label{app:np-perturbations}

We derive the radiative equations for positive and negative spin weights
for coupled gravitational and electromagnetic perturbations of
Kerr--Newman--de Sitter at finite $\Lambda$.
The resulting closed first-order systems,
Eqs.~\eqref{eq:bg-positive-system} and \eqref{eq:bg-negative-system},
provide the starting point for the chiral transformation in
Sec.~\ref{sec:chiral-system}.

Following the phantom-gauge reduction for Kerr--Newman
perturbations described in
Sec.~111 in Chap.~11 of Ref.~\cite{Chandrasekhar:1983},
we first fix the perturbed tetrad. We then linearize the relevant
Bianchi and Ricci identities and use the Maxwell equations to
close the system for positive spin weights. We write this system for Fourier
modes and obtain the counterpart for negative spin weights by the tetrad
prime transformation.

\subsubsection{Linear perturbations and phantom gauge}

A superscript $(1)$ denotes a linear perturbation about the
Kerr--Newman--de Sitter background.
We impose the phantom gauge by two successive infinitesimal
null rotations of the perturbed tetrad. Their complex parameters
$\zeta_k,\zeta_l$ are of first perturbative order, and all
transformations below are written to that order.
Both rotations preserve the normalization and orthogonality
of the null tetrad at this order.

First, hold $\bm{k}_{\mathrm{NP}}$ fixed and apply
\begin{equation}
 \begin{aligned}
 \bm{k}_{\mathrm{NP}}&\mapsto\bm{k}_{\mathrm{NP}},&
 \bm{l}_{\mathrm{NP}}&\mapsto\bm{l}_{\mathrm{NP}}
       +\bar\zeta_k\bm{m}_{\mathrm{NP}}
       +\zeta_k\overline{\bm{m}}_{\mathrm{NP}},\\
 \bm{m}_{\mathrm{NP}}&\mapsto\bm{m}_{\mathrm{NP}}
       +\zeta_k\bm{k}_{\mathrm{NP}},&
 \overline{\bm{m}}_{\mathrm{NP}}&\mapsto\overline{\bm{m}}_{\mathrm{NP}}
       +\bar\zeta_k\bm{k}_{\mathrm{NP}}.
 \end{aligned}
 \label{eq:np-null-rotation-k}
\end{equation}
Since $\Phi_0=\Phi_2=0$ on the background, the definitions
in Eq.~\eqref{eq:np-maxwell-definitions} give
\begin{equation}
 \Phi_2^{(1)}
 \mapsto\Phi_2^{(1)}+2\bar\zeta_k\Phi_1,
 \qquad
 \Phi_0^{(1)}
 \mapsto\Phi_0^{(1)}.
 \label{eq:np-maxwell-null-rotations}
\end{equation}
Here $\Phi_1=Q/(2w^2)\ne0$ is the background value.
Choosing $\bar\zeta_k=-\Phi_2^{(1)}/(2\Phi_1)$
sets $\Phi_2^{(1)}=0$ while leaving
$\Phi_0^{(1)}$ unchanged.

Next, hold the resulting $\bm{l}_{\mathrm{NP}}$ fixed and apply
\begin{equation}
 \begin{aligned}
 \bm{l}_{\mathrm{NP}}&\mapsto\bm{l}_{\mathrm{NP}},&
 \bm{k}_{\mathrm{NP}}&\mapsto\bm{k}_{\mathrm{NP}}
       +\bar\zeta_l\bm{m}_{\mathrm{NP}}
       +\zeta_l\overline{\bm{m}}_{\mathrm{NP}},\\
 \bm{m}_{\mathrm{NP}}&\mapsto\bm{m}_{\mathrm{NP}}
       +\zeta_l\bm{l}_{\mathrm{NP}},&
 \overline{\bm{m}}_{\mathrm{NP}}&\mapsto\overline{\bm{m}}_{\mathrm{NP}}
       +\bar\zeta_l\bm{l}_{\mathrm{NP}}.
 \end{aligned}
 \label{eq:np-null-rotation-l}
\end{equation}
The Maxwell perturbations now transform as
\begin{equation}
 \Phi_0^{(1)}
 \mapsto\Phi_0^{(1)}+2\zeta_l\Phi_1,
 \qquad
 \Phi_2^{(1)}
 \mapsto\Phi_2^{(1)}=0.
 \label{eq:np-maxwell-l-rotation}
\end{equation}
The first condition is preserved because
$\Phi_2=F_{\bar m l}$ is unchanged when
$\overline{\bm{m}}_{\mathrm{NP}}$ is shifted by a multiple of
$\bm{l}_{\mathrm{NP}}$, using $F_{ll}=0$.
Choosing $\zeta_l=-\Phi_0^{(1)}/(2\Phi_1)$
then gives the phantom gauge locally at linear order:
\begin{equation}
 \Phi_0^{(1)}=
 \Phi_2^{(1)}=0.
 \label{eq:bg-phantom}
\end{equation}
This is a choice of perturbed tetrad; it does not set
the electromagnetic perturbation to zero.

Using the phantom-gauge conditions in the Einstein--Maxwell relation
$\Phi_{ij}=2\Phi_i\bar\Phi_j$, we find that all Ricci scalars
$\Phi_{ij}$ except $\Phi_{11}$ have vanishing linear parts.
The perturbation $\Phi_{11}^{(1)}$ need not vanish.

\subsubsection{Newman--Penrose equations for positive spin weights}

A field has spin weight $s$ if it acquires a factor $e^{is\vartheta}$
under the tetrad rotation
$\bm{m}_{\mathrm{NP}}\mapsto e^{i\vartheta}\bm{m}_{\mathrm{NP}}$,
with $\bm{k}_{\mathrm{NP}},\bm{l}_{\mathrm{NP}}$ fixed.
We first derive the equations for positive spin weights.
The equations for negative spin weights then follow by interchanging
the null tetrad directions, as described below.

\paragraph{Bianchi and Ricci identities.}

We first use two Bianchi identities and one Ricci identity\linebreak
(Eqs.~(7.32a), (7.32b), and (7.21b) of Ref.~\cite{Stephani:2003}).
All quantities in these identities refer to the full fields,
before linearization:
\begin{align}
 &(\bar\delta-4\alpha_{\mathrm{NP}}+\pi_{\mathrm{NP}})\Psi_0
 -(D-4\rho_{\mathrm{NP}}-2\epsilon_{\mathrm{NP}})\Psi_1-3\kappa_{\mathrm{NP}}\Psi_2
 \nonumber\\
 &~=(\delta+\bar\pi_{\mathrm{NP}}-2\bar\alpha_{\mathrm{NP}}-2\beta_{\mathrm{NP}})\Phi_{00}
 -(D-2\epsilon_{\mathrm{NP}}-2\bar\rho_{\mathrm{NP}})\Phi_{01}
% \nonumber\\ &\qquad\quad
+2\sigma_{\mathrm{NP}}\Phi_{10}-2\kappa_{\mathrm{NP}}\Phi_{11}-\bar\kappa_{\mathrm{NP}}\Phi_{02},
 \label{eq:np-full-bianchi}\\
 &(\Delta-4\gamma_{\mathrm{NP}}+\mu_{\mathrm{NP}})\Psi_0
 -(\delta-4\tau_{\mathrm{NP}}-2\beta_{\mathrm{NP}})\Psi_1-3\sigma_{\mathrm{NP}}\Psi_2
 \nonumber\\
 &~=(\delta+2\bar\pi_{\mathrm{NP}}-2\beta_{\mathrm{NP}})\Phi_{01}
 -(D-2\epsilon_{\mathrm{NP}}+2\bar\epsilon_{\mathrm{NP}}-\bar\rho_{\mathrm{NP}})\Phi_{02}
% \nonumber\\ &\qquad\quad
-\bar\lambda_{\mathrm{NP}}\Phi_{00}+2\sigma_{\mathrm{NP}}\Phi_{11}-2\kappa_{\mathrm{NP}}\Phi_{12},
 \label{eq:np-second-ricci-side}\\
 &D\sigma_{\mathrm{NP}}-\delta\kappa_{\mathrm{NP}}
 =(\rho_{\mathrm{NP}}+\bar\rho_{\mathrm{NP}}+3\epsilon_{\mathrm{NP}}-\bar\epsilon_{\mathrm{NP}})\sigma_{\mathrm{NP}}
 %\nonumber\\ &\qquad\quad
 -(\tau_{\mathrm{NP}}-\bar\pi_{\mathrm{NP}}+\bar\alpha_{\mathrm{NP}}+3\beta_{\mathrm{NP}})\kappa_{\mathrm{NP}}+\Psi_0.
 \label{eq:np-full-ricci}
\end{align}

\begingroup
\emergencystretch=1em
Linearizing Eqs.~\eqref{eq:np-full-bianchi}--\eqref{eq:np-full-ricci}
about the Kerr--Newman--de Sitter background, with the
Einstein--Maxwell equations and phantom gauge imposed, we obtain
\par\endgroup
\begin{align}
 (\bar\delta-4\alpha_{\mathrm{NP}}+\pi_{\mathrm{NP}})\Psi_0^{(1)}
 -(D-4\rho_{\mathrm{NP}}-2\epsilon_{\mathrm{NP}})\Psi_1^{(1)}&=(3\Psi_2-2\Phi_{11})\kappa_{\mathrm{NP}}^{(1)},
 \label{eq:np-bianchi1}\\
 (\Delta-4\gamma_{\mathrm{NP}}+\mu_{\mathrm{NP}})\Psi_0^{(1)}
 -(\delta-4\tau_{\mathrm{NP}}-2\beta_{\mathrm{NP}})\Psi_1^{(1)}&=(3\Psi_2+2\Phi_{11})\sigma_{\mathrm{NP}}^{(1)},
 \label{eq:np-bianchi2}\\
 \begin{aligned}
 &(D-\rho_{\mathrm{NP}}-\bar\rho_{\mathrm{NP}}-3\epsilon_{\mathrm{NP}}+\bar\epsilon_{\mathrm{NP}})\sigma_{\mathrm{NP}}^{(1)}\\
 &\quad -(\delta-\tau_{\mathrm{NP}}+\bar\pi_{\mathrm{NP}}-\bar\alpha_{\mathrm{NP}}-3\beta_{\mathrm{NP}})\kappa_{\mathrm{NP}}^{(1)}
 \end{aligned}&=\Psi_0^{(1)}.
 \label{eq:np-ricci}
\end{align}
Here all coefficients and differential operators are evaluated
on the background.
Although $\epsilon_{\mathrm{NP}}=0$ in the background tetrad,
we retain it formally here to obtain the equations for negative spin
weights by the prime transformation below.

\paragraph{Maxwell compatibility.}

The fourth equation follows from the Maxwell compatibility condition
and two Ricci identities.
The source-free Maxwell equations have the Newman--Penrose components
(Eqs.~(7.22)--(7.25) of Ref.~\cite{Stephani:2003})
\begin{equation}
 \begin{aligned}
 D\Phi_1-\bar\delta\Phi_0
 &=(\pi_{\mathrm{NP}}-2\alpha_{\mathrm{NP}})\Phi_0
   +2\rho_{\mathrm{NP}}\Phi_1-\kappa_{\mathrm{NP}}\Phi_2,\\
 \delta\Phi_1-\Delta\Phi_0
 &=(\mu_{\mathrm{NP}}-2\gamma_{\mathrm{NP}})\Phi_0
   +2\tau_{\mathrm{NP}}\Phi_1-\sigma_{\mathrm{NP}}\Phi_2,\\
 D\Phi_2-\bar\delta\Phi_1
 &=-\lambda_{\mathrm{NP}}\Phi_0+2\pi_{\mathrm{NP}}\Phi_1
   +(\rho_{\mathrm{NP}}-2\epsilon_{\mathrm{NP}})\Phi_2,\\
 \delta\Phi_2-\Delta\Phi_1
 &=-\nu_{\mathrm{NP}}\Phi_0+2\mu_{\mathrm{NP}}\Phi_1
   +(\tau_{\mathrm{NP}}-2\beta_{\mathrm{NP}})\Phi_2.
 \end{aligned}
 \label{eq:np-full-maxwell}
\end{equation}

Imposing the phantom gauge and retaining terms through first
order gives
\begin{equation}
 \begin{gathered}
 D\Phi_1=2\rho_{\mathrm{NP}}\Phi_1,\qquad
 \delta\Phi_1=2\tau_{\mathrm{NP}}\Phi_1,\\
 \Delta\Phi_1=-2\mu_{\mathrm{NP}}\Phi_1,\qquad
 \bar\delta\Phi_1=-2\pi_{\mathrm{NP}}\Phi_1.
 \end{gathered}
 \label{eq:np-maxwell-two}
\end{equation}
Here $\Phi_1$, the directional derivatives, and the spin
coefficients include their linear variations.

We also use the scalar
commutator (Eq.~(7.6b) of Ref.~\cite{Stephani:2003})
\begin{equation}
 [\delta,D]=(\bar\alpha_{\mathrm{NP}}+\beta_{\mathrm{NP}}-\bar\pi_{\mathrm{NP}})D+\kappa_{\mathrm{NP}}\Delta
       -(\bar\rho_{\mathrm{NP}}+\epsilon_{\mathrm{NP}}-\bar\epsilon_{\mathrm{NP}})\delta-\sigma_{\mathrm{NP}}\bar\delta.
 \label{eq:np-commutator}
\end{equation}
Substituting Eq.~\eqref{eq:np-maxwell-two} into the scalar commutator
in Eq.~\eqref{eq:np-commutator} gives a compatibility condition.
Since $[\delta,D]\Phi_1=2\Phi_1(\delta\rho_{\mathrm{NP}}-D\tau_{\mathrm{NP}})$
and the background $\Phi_1$ is nonzero, division by $2\Phi_1$ yields
\begin{align}
 \delta\rho_{\mathrm{NP}}-D\tau_{\mathrm{NP}}
 ={}&(\bar\alpha_{\mathrm{NP}}+\beta_{\mathrm{NP}}-\bar\pi_{\mathrm{NP}})\rho_{\mathrm{NP}}
 -(\bar\rho_{\mathrm{NP}}+\epsilon_{\mathrm{NP}}-\bar\epsilon_{\mathrm{NP}})\tau_{\mathrm{NP}}
 -\kappa_{\mathrm{NP}}\mu_{\mathrm{NP}}+\sigma_{\mathrm{NP}}\pi_{\mathrm{NP}}.
 \label{eq:np-maxwell-compatibility}
\end{align}

We also use the following Ricci identities
(Eqs.~(7.21c) and (7.21k) of Ref.~\cite{Stephani:2003}):
\begin{align}
 D\tau_{\mathrm{NP}}-\Delta\kappa_{\mathrm{NP}}
 ={}&\rho_{\mathrm{NP}}(\tau_{\mathrm{NP}}+\bar\pi_{\mathrm{NP}})
 +\sigma_{\mathrm{NP}}(\bar\tau_{\mathrm{NP}}+\pi_{\mathrm{NP}})
 \nonumber\\
 &+\tau_{\mathrm{NP}}(\epsilon_{\mathrm{NP}}-\bar\epsilon_{\mathrm{NP}})
 -\kappa_{\mathrm{NP}}(3\gamma_{\mathrm{NP}}+\bar\gamma_{\mathrm{NP}})
 +\Psi_1+\Phi_{01},\nonumber\\
 \delta\rho_{\mathrm{NP}}-\bar\delta\sigma_{\mathrm{NP}}
 ={}&\rho_{\mathrm{NP}}(\bar\alpha_{\mathrm{NP}}+\beta_{\mathrm{NP}})
 -\sigma_{\mathrm{NP}}(3\alpha_{\mathrm{NP}}-\bar\beta_{\mathrm{NP}})
 \nonumber\\
 &+\tau_{\mathrm{NP}}(\rho_{\mathrm{NP}}-\bar\rho_{\mathrm{NP}})
 +\kappa_{\mathrm{NP}}(\mu_{\mathrm{NP}}-\bar\mu_{\mathrm{NP}})
 -\Psi_1+\Phi_{01}.
 \label{eq:np-compatibility-identities}
\end{align}

Subtracting the two identities in
Eq.~\eqref{eq:np-compatibility-identities} and using
Eq.~\eqref{eq:np-maxwell-compatibility} to eliminate
$\delta\rho_{\mathrm{NP}}-D\tau_{\mathrm{NP}}$,
we obtain at linear order
\begin{equation}
 (\Delta-3\gamma_{\mathrm{NP}}-\bar\gamma_{\mathrm{NP}}-2\mu_{\mathrm{NP}}+\bar\mu_{\mathrm{NP}})\kappa_{\mathrm{NP}}^{(1)}
 -(\bar\delta-3\alpha_{\mathrm{NP}}+\bar\beta_{\mathrm{NP}}-\bar\tau_{\mathrm{NP}}-2\pi_{\mathrm{NP}})\sigma_{\mathrm{NP}}^{(1)}=-2\Psi_1^{(1)}.
 \label{eq:np-fourth}
\end{equation}
All coefficients and differential operators here are evaluated
on the background.

\subsubsection{Explicit form on Kerr--Newman--de Sitter}

We now substitute the Kerr--Newman--de Sitter background quantities
into the Newman--Penrose equations \eqref{eq:np-bianchi1}--\eqref{eq:np-ricci}
and \eqref{eq:np-fourth} and express them in terms of Fourier modes
and rescaled perturbation variables.
We consider modes proportional to $e^{-i\omega t+im\phi}$, with
$m\in\mathbb Z$ and initially $\omega\in\mathbb C$.
We collect the frequency and azimuthal dependence in the functions
\begin{equation}
 K(r)\coloneqq \Xi\bigl(-(r^2+a^2)\omega+am\bigr),\qquad
 W(\theta)\coloneqq \frac{\Xi(m-a\omega\sin^2\theta)}{h\sin\theta}.
 \label{eq:bg-mode-potentials}
\end{equation}
Using these functions, we define the radial operators
\begin{align}
 \mathcal D_j&\coloneqq \partial_r+\frac{iK}{\Delta_r}+j\frac{\Delta_r'}{\Delta_r},
 &\qquad \mathcal D_j^\dagger&\coloneqq \partial_r-\frac{iK}{\Delta_r}+j\frac{\Delta_r'}{\Delta_r},
 \label{eq:bg-mode-operators}
\end{align}
and the angular operators
\begin{equation}
 \begin{aligned}
 \mathcal L_j&\coloneqq h\partial_\theta+W+j(\partial_\theta h+h\cot\theta),\\
 \mathcal L_j^\dagger&\coloneqq h\partial_\theta-W+j(\partial_\theta h+h\cot\theta).
 \end{aligned}
 \label{eq:bg-angular-mode-operators}
\end{equation}
The dagger labels the displayed sign reversal
at fixed $\omega$ for the differential operators and denotes the
conjugate transpose for finite matrices.

To abbreviate the background coefficients in
Eqs.~\eqref{eq:np-bianchi1} and \eqref{eq:np-bianchi2}, we introduce
\begin{equation}
 \mathcal F_\pm\coloneqq 3\left(M-\frac{Q^2}z\right)\pm\frac{Q^2w}{z^2}.
 \label{eq:bg-Fpm}
\end{equation}

Substituting the tetrad in Eqs.~\eqref{eq:bg-tetrad-k}--\eqref{eq:bg-tetrad}
into the definitions of the directional derivatives gives, on these modes,
\begin{equation}
 D=\mathcal D_0,\qquad
 \Delta=-\frac{\Delta_r}{2\Sigma}\mathcal D_0^\dagger,\qquad
 \delta=\frac{\mathcal L_0^\dagger}{\sqrt2z},\qquad
 \bar\delta=\frac{\mathcal L_0}{\sqrt2w}.
 \label{eq:np-directional-derivatives}
\end{equation}
To obtain the coordinate equations, substitute Eqs.~\eqref{eq:np-background-coefficients} and \eqref{eq:np-directional-derivatives}, and use
\begin{equation}
 w^3(3\Psi_2-4|\Phi_1|^2)=-\mathcal F_+,
 \qquad
 w^3(3\Psi_2+4|\Phi_1|^2)=-\mathcal F_-.
 \label{eq:np-F-identities}
\end{equation}
Following the normalizations of Ref.~\cite{Hintz:2026jiw}, define
the following variables, with hats denoting normalized first-order
perturbations of the corresponding Newman--Penrose quantities:
\begin{equation}
 \widehat{\Psi}_0\coloneqq \Psi_0^{(1)},\qquad
 \widehat{\Psi}_1\coloneqq \sqrt2w\,\Psi_1^{(1)},\qquad
 \widehat{\kappa}\coloneqq \frac{\kappa_{\mathrm{NP}}^{(1)}}{\sqrt2w^2},\qquad
 \widehat{\sigma}\coloneqq \frac{z\,\sigma_{\mathrm{NP}}^{(1)}}{w^2}.
 \label{eq:bg-positive-fields}
\end{equation}
The spin weights of $(\widehat{\Psi}_0,\widehat{\Psi}_1,\widehat{\kappa},\widehat{\sigma})$ are $(2,1,1,2)$.
Substituting these definitions and the background Newman--Penrose quantities into
Eqs.~\eqref{eq:np-bianchi1}, \eqref{eq:np-bianchi2},
\eqref{eq:np-ricci}, and \eqref{eq:np-fourth} gives, in the same order,
\begin{subequations}\label{eq:bg-positive-system}
\begin{align}
 (\mathcal L_2-3iv/w)\widehat{\Psi}_0
 -(\mathcal D_0+3/w)\widehat{\Psi}_1&=-2\mathcal F_+ \widehat{\kappa},
 \label{eq:bg-positive-a}\\
 \Delta_r(\mathcal D_2^\dagger-3/w)\widehat{\Psi}_0
 +(\mathcal L_{-1}^\dagger+3iv/w)\widehat{\Psi}_1&=2\mathcal F_- \widehat{\sigma},
 \label{eq:bg-positive-b}\\
 (\mathcal D_0+3/w)\widehat{\sigma}
 -(\mathcal L_{-1}^\dagger+3iv/w)\widehat{\kappa}&=\frac z{w^2}\widehat{\Psi}_0,
 \label{eq:bg-positive-c}\\
 \Delta_r(\mathcal D_2^\dagger-3/w)\widehat{\kappa}
 +(\mathcal L_2-3iv/w)\widehat{\sigma}&=\frac{2z}{w^2}\widehat{\Psi}_1.
 \label{eq:bg-positive-d}
\end{align}
\end{subequations}

\subsubsection{Equations for negative spin weights}

The tetrad prime used here is the algebraic exchange
\begin{equation}
 (\bm{k}_{\mathrm{NP}},\bm{l}_{\mathrm{NP}},\bm{m}_{\mathrm{NP}},\overline{\bm{m}}_{\mathrm{NP}})'=(\bm{l}_{\mathrm{NP}},\bm{k}_{\mathrm{NP}},\overline{\bm{m}}_{\mathrm{NP}},\bm{m}_{\mathrm{NP}}).
 \label{eq:np-prime}
\end{equation}
The directional derivatives transform as
\begin{equation}
 (D,\Delta,\delta,\bar\delta)'
 =(\Delta,D,\bar\delta,\delta).
\end{equation}
The coefficient transformations are
\begin{equation}
 \kappa_{\mathrm{NP}}'=-\nu_{\mathrm{NP}},\quad \sigma_{\mathrm{NP}}'=-\lambda_{\mathrm{NP}},\quad
 \rho_{\mathrm{NP}}'=-\mu_{\mathrm{NP}},\quad\tau_{\mathrm{NP}}'=-\pi_{\mathrm{NP}},\quad\epsilon_{\mathrm{NP}}'=-\gamma_{\mathrm{NP}},\quad\beta_{\mathrm{NP}}'=-\alpha_{\mathrm{NP}},
 \label{eq:np-prime-coefficients}
\end{equation}
and their inverse relations. Weyl scalars obey $\Psi_j'=\Psi_{4-j}$, while
$(\Phi_0',\Phi_1',\Phi_2')=(-\Phi_2,-\Phi_1,-\Phi_0)$.
Both the phantom gauge and $\Phi_{11}$ are preserved.
Under this transformation, $\epsilon_{\mathrm{NP}}'=-\gamma_{\mathrm{NP}}$,
which need not vanish.
We must therefore apply the prime transformation to the identities before
setting $\epsilon_{\mathrm{NP}}=0$; otherwise, the terms arising from
$\epsilon_{\mathrm{NP}}'=-\gamma_{\mathrm{NP}}$ would be lost.

The linear perturbations transform as
\begin{equation}
 (\Psi_0^{(1)},\Psi_1^{(1)},\kappa_{\mathrm{NP}}^{(1)},\sigma_{\mathrm{NP}}^{(1)})'=(\Psi_4^{(1)},\Psi_3^{(1)},-\nu_{\mathrm{NP}}^{(1)},-\lambda_{\mathrm{NP}}^{(1)}).
\end{equation}
Applying the prime transformation to
Eqs.~\eqref{eq:np-bianchi1}, \eqref{eq:np-bianchi2},
\eqref{eq:np-ricci}, and \eqref{eq:np-fourth},
and then setting the background $\epsilon_{\mathrm{NP}}=0$,
gives the following equations in the original background
frame, in the same order.
\begin{gather}
 (\delta+4\beta_{\mathrm{NP}}-\tau_{\mathrm{NP}})\Psi_4^{(1)}-(\Delta+4\mu_{\mathrm{NP}}+2\gamma_{\mathrm{NP}})\Psi_3^{(1)}
 =-(3\Psi_2-4|\Phi_1|^2)\nu_{\mathrm{NP}}^{(1)},
 \label{eq:np-negative-bianchi1}\\
 (D-\rho_{\mathrm{NP}})\Psi_4^{(1)}-(\bar\delta+4\pi_{\mathrm{NP}}+2\alpha_{\mathrm{NP}})\Psi_3^{(1)}
 =-(3\Psi_2+4|\Phi_1|^2)\lambda_{\mathrm{NP}}^{(1)},
 \label{eq:np-negative-bianchi2}\\
 (\bar\delta+\pi_{\mathrm{NP}}-\bar\tau_{\mathrm{NP}}+\bar\beta_{\mathrm{NP}}+3\alpha_{\mathrm{NP}})\nu_{\mathrm{NP}}^{(1)}
 -(\Delta+\mu_{\mathrm{NP}}+\bar\mu_{\mathrm{NP}}+3\gamma_{\mathrm{NP}}-\bar\gamma_{\mathrm{NP}})\lambda_{\mathrm{NP}}^{(1)}=\Psi_4^{(1)},
 \label{eq:np-negative-ricci}\\
 (D+2\rho_{\mathrm{NP}}-\bar\rho_{\mathrm{NP}})\nu_{\mathrm{NP}}^{(1)}
 -(\delta+3\beta_{\mathrm{NP}}-\bar\alpha_{\mathrm{NP}}+\bar\pi_{\mathrm{NP}}+2\tau_{\mathrm{NP}})\lambda_{\mathrm{NP}}^{(1)}=2\Psi_3^{(1)}.
 \label{eq:np-negative-four}
\end{gather}
For the system for negative spin weights, define
\begin{equation}
 \widehat{\Psi}_4\coloneqq w^4\Psi_4^{(1)},\qquad
 \widehat{\Psi}_3\coloneqq \frac{w^3}{\sqrt2}\Psi_3^{(1)},\qquad
 \widehat{\nu} \coloneqq \frac{zw}{\sqrt2}\nu_{\mathrm{NP}}^{(1)},\qquad
 \widehat{\lambda} \coloneqq \frac w2\lambda_{\mathrm{NP}}^{(1)}.
 \label{eq:bg-negative-fields}
\end{equation}
The spin weights of $(\widehat{\Psi}_4,\widehat{\Psi}_3,\widehat{\nu},\widehat{\lambda})$
are $(-2,-1,-1,-2)$.
Substituting these definitions and the background Newman--Penrose quantities into
Eqs.~\eqref{eq:np-negative-bianchi1}--\eqref{eq:np-negative-four} gives
\begin{subequations}\label{eq:bg-negative-system}
\begin{align}
 (\mathcal L_2^\dagger-3iv/w)\widehat{\Psi}_4
 +\Delta_r(\mathcal D_{-1}^\dagger+3/w)\widehat{\Psi}_3&=2\mathcal F_+ \widehat{\nu},
 \label{eq:bg-negative-a}\\
 (\mathcal D_0-3/w)\widehat{\Psi}_4
 -(\mathcal L_{-1}+3iv/w)\widehat{\Psi}_3&=2\mathcal F_- \widehat{\lambda},
 \label{eq:bg-negative-b}\\
 \Delta_r(\mathcal D_{-1}^\dagger+3/w)\widehat{\lambda}
 +(\mathcal L_{-1}+3iv/w)\widehat{\nu}&=\frac z{w^2}\widehat{\Psi}_4,
 \label{eq:bg-negative-c}\\
 (\mathcal D_0-3/w)\widehat{\nu}
 -(\mathcal L_2^\dagger-3iv/w)\widehat{\lambda}&=\frac{2z}{w^2}\widehat{\Psi}_3.
 \label{eq:bg-negative-d}
\end{align}
\end{subequations}
Each system closes on four radiative variables and is a necessary
subsystem of the linearized Einstein--Maxwell equations.
Our goal is to separate the radial and angular dependence of the coupled
systems \eqref{eq:bg-positive-system} and \eqref{eq:bg-negative-system}.

\subsection{Chiral form}
\label{sec:chiral-system}

Following the transformation introduced for Kerr--Newman
perturbations by Hintz~\cite{Hintz:2026jiw}, we rewrite the Kerr--Newman--de Sitter systems
in a common matrix form.

We use sans-serif Latin letters for finite matrices, while keeping
the conventional notation for the Pauli matrices.
For each sign of the original spin weights, we combine the four
perturbation variables into two-component fields $y_+^{(\varsigma)}$
and $y_-^{(\varsigma)}$, whose equations have principal derivatives
$u\partial_r-ih\partial_\theta$ and $u\partial_r+ih\partial_\theta$,
respectively. We call this arrangement the chiral form.
The subscripts $+$ and $-$ label the derivative combinations,
while $\varsigma=+1$ and $-1$ specify the systems for positive
and negative spin weights. We define
\begin{align}
 y_\pm^{(+)}&\coloneqq \Delta_r^{5/4}\sqrt{h\sin\theta}
 \begin{pmatrix}\widehat{\Psi}_0\pm i\widehat{\Psi}_1/u\\-2Q(\widehat{\kappa}\pm i \widehat{\sigma}/u)\end{pmatrix},
 \label{eq:bg-chiral-plus}\\
 y_\pm^{(-)}&\coloneqq \Delta_r^{-3/4}\sqrt{h\sin\theta}
 \begin{pmatrix}\widehat{\Psi}_4\mp iu\widehat{\Psi}_3\\2Q(\widehat{\nu}\mp iu\widehat{\lambda})\end{pmatrix}.
 \label{eq:bg-chiral-minus}
\end{align}
The principal derivative matrices in Eqs.~\eqref{eq:bg-positive-system} and \eqref{eq:bg-negative-system} have eigenvectors with relative coefficient $\pm i$. The factors $\Delta_r^{5/4}$ and $\Delta_r^{-3/4}$, together with $\sqrt{h\sin\theta}$, collect the scalar derivative terms. Direct substitution gives
\begin{equation}
 (\mathscr D-\mathsf{A})y_+^{(\varsigma)}=\mathsf{F}^{(\varsigma)}y_-^{(\varsigma)},
 \qquad
 (\overline{\mathscr D}+\mathsf{A})y_-^{(\varsigma)}
 =\operatorname{adj}\mathsf{F}^{(\varsigma)}y_+^{(\varsigma)}.
 \label{eq:bg-chiral-system}
\end{equation}
The differential operators are defined by
\begin{equation}
 \mathscr D\coloneqq u\partial_r-ih\partial_\theta,
 \qquad
 \overline{\mathscr D}\coloneqq u\partial_r+ih\partial_\theta.
 \label{eq:bg-chiral-derivatives}
\end{equation}
The matrix $\mathsf A$ is given by
\begin{equation}
 \mathsf{A}\coloneqq
 \begin{pmatrix}
 0&-iQw/z^2\\
 iQz/w^2&0
 \end{pmatrix}.
 \label{eq:bg-chiral-matrices}
\end{equation}
The coupling matrix $\mathsf F^{(\varsigma)}$ is written as
\begin{equation}
 \mathsf{F}^{(\varsigma)}
 \coloneqq c^{(\varsigma)}\mathsf{I}-\mathsf{B}^{(\varsigma)}.
 \label{eq:bg-chiral-F}
\end{equation}
Here $\mathsf B^{(\varsigma)}$ is the traceless matrix
\begin{equation}
 \mathsf{B}^{(\varsigma)}\coloneqq i
 \begin{pmatrix}
 b^{(\varsigma)}&3(M-Q^2/z)/Q\\
 3Qz/w^2&-b^{(\varsigma)}
 \end{pmatrix},
 \label{eq:bg-chiral-B}
\end{equation}
where the scalar coefficients are
\begin{equation}
 \begin{aligned}
 b^{(\varsigma)}&\coloneqq -\varsigma W-\frac32(\partial_\theta h+h\cot\theta)+\frac{3iv}w,\\
 c^{(\varsigma)}&\coloneqq \varsigma\frac{iK}u+\frac{3u}w-\frac{3\Delta_r'}{4u}.
 \end{aligned}
 \label{eq:bg-chiral-bc}
\end{equation}
Here $\operatorname{adj}\mathsf{M}=(\operatorname{tr}\mathsf{M})\mathsf{I}-\mathsf{M}$ for a $2\times2$ matrix. The bar on $\overline{\mathscr D}$ changes $i$ to $-i$ in the differential operator, without conjugating the mode frequency. Since $\mathsf{B}^{(\varsigma)}$ is traceless, the reverse coupling is exactly
$c^{(\varsigma)}\mathsf{I}+\mathsf{B}^{(\varsigma)}
=\operatorname{adj}(c^{(\varsigma)}\mathsf{I}-\mathsf{B}^{(\varsigma)})$.

The off-diagonal terms give the same $\mathsf A$ in the systems for both signs of spin weight,
while $(K,W)$ change sign. The matrix $\mathsf{A}$ is Hermitian
and independent of $\omega,m,\varsigma$. The coupling matrix satisfies
\begin{equation}
 \mathsf{F}^{(-)}(\omega,m)=\mathsf{F}^{(+)}(-\omega,-m).
 \label{eq:bg-source-affine}
\end{equation}
This identity will allow the same transformation to separate
the systems for both signs of spin weight.

Equation~\eqref{eq:bg-chiral-system} has the same matrix
structure as the Kerr--Newman system in Eq.~(4) of
Ref.~\cite{Hintz:2026jiw}.
The matrix $\mathsf A$ is unchanged, while the cosmological
constant enters the differential operators and
$\mathsf F^{(\varsigma)}$ through $\Delta_r$,
$h=\sqrt{\Delta_\theta}$, and $\Xi$.
Setting $\Lambda=0$ recovers Hintz's system.

\section{The separation transformation and its properties}
\label{sec:transformation}

In this section, we construct an invertible matrix transformation
$\mathsf X(r,\theta)$ that separates the chiral system
\eqref{eq:bg-chiral-system} into two-component angular and radial
systems linked by a common separation constant.
The same transformation applies to both signs of spin weight.
The central task is to construct $\mathsf X$ and verify the
properties that make this separation possible.

Sec.~\ref{sec:target-transformation} first identifies sufficient
conditions on $\mathsf X$ and derives the separated equations
under these conditions.
Sec.~\ref{sec:main-result} then states the main results.
We define $\mathsf X$ explicitly in
Sec.~\ref{sec:explicit-transformation} and verify its separation
properties and boundary regularity in
Sec.~\ref{sec:verification-transformation}.
The explicit coefficient functions are obtained in
Secs.~\ref{sec:integrating-factor} and \ref{sec:source-normalization}.
Finally, Sec.~\ref{sec:local-kn-limit} examines the local
Kerr--Newman limit and relates our transformation to Hintz's
matrix integrating factor.

\subsection{Matrix structure for separation of variables}
\label{sec:target-transformation}
\label{sec:separation-mechanism}
\label{sec:paired-equations}

\paragraph{Conditions on the transformation.}

The differential operators in the chiral system are
$u(r)\partial_r\mp ih(\theta)\partial_\theta$.
The coefficients of the radial and angular derivatives
depend only on $r$ and $\theta$, respectively.
We seek a change of variables that also separates the terms involving $\mathsf A$ and
the coupling.
The transformed derivative equations need not be free of
matrix terms. We allow a residual that can be incorporated
into the radial and angular equations separately.
We use the Pauli matrices
\begin{equation}
 \sigma_1\coloneqq \begin{pmatrix}0&1\\1&0\end{pmatrix},\qquad
 \sigma_2\coloneqq \begin{pmatrix}0&-i\\i&0\end{pmatrix},\qquad
 \sigma_3\coloneqq \begin{pmatrix}1&0\\0&-1\end{pmatrix},
 \label{eq:sep-pauli}
\end{equation}
and write $\mathsf I\coloneqq \mathsf I_2$.
Suppose an invertible matrix $\mathsf X(r,\theta)$, independent of
$\omega,m,\varsigma$, satisfies
\begin{equation}
 \det\mathsf X=1,\qquad
 \mathscr D\mathsf X=\mathsf A\mathsf X
                   +\mathsf X\bigl(\alpha_r(r)+\alpha_\theta(\theta)\bigr)\sigma_2.
 \label{eq:main-X}
\end{equation}
Here $\alpha_r,\alpha_\theta$ are real functions depending only on the background.
Multiplication of $\mathsf X$ by a nonzero complex constant
preserves separability, with a corresponding constant
rescaling of the transformed variables.
We impose $\det\mathsf X=1$ as an overall normalization convention.
Further freedom in $\mathsf X$ that preserves separation is
discussed in Sec.~\ref{sec:local-kn-limit}.
For Kerr--Newman, Hintz's matrix integrating
factor~\cite{Hintz:2026jiw}, normalized to unit determinant,
corresponds to
$\alpha_r=\alpha_\theta=0$ in Eq.~\eqref{eq:main-X}.
Here we allow the separated term
$\bigl(\alpha_r(r)+\alpha_\theta(\theta)\bigr)\sigma_2$;
the matrix constructed in Sec.~\ref{sec:explicit-transformation}
satisfies Eq.~\eqref{eq:main-X} for Kerr--Newman--de Sitter.

Define
\begin{equation}
 \mathsf C^{(\varsigma)}
 \coloneqq -\mathsf X^{-1}\mathsf F^{(\varsigma)}
                         (\mathsf X^\dagger)^{-1}.
 \label{eq:main-C-definition}
\end{equation}
The matrix $\mathsf C^{(\varsigma)}$ contains no derivatives of the fields.
The remaining requirement for $\mathsf X$ is
\begin{equation}
 \mathsf C^{(\varsigma)}
 =C_0^{(\varsigma)}(r)\mathsf I+C_2^{(\varsigma)}(r)\sigma_2
   +i C_1^{(\varsigma)}(\theta)\sigma_1
   +i C_3^{(\varsigma)}(\theta)\sigma_3.
 \label{eq:main-separated-C}
\end{equation}
These four coefficients may depend on $\omega,m,\varsigma$;
their arguments specify the coordinate dependence.
Equations~\eqref{eq:main-X} and \eqref{eq:main-separated-C}
are sufficient: the coefficient $\alpha_r(r)+\alpha_\theta(\theta)$ can be included
in the radial and angular equations separately.
The transformation is constructed in Sec.~\ref{sec:explicit-transformation};
its coefficient functions are obtained in
Secs.~\ref{sec:integrating-factor} and \ref{sec:source-normalization}.

\paragraph{Derivation of the separated equations.}

Define
\begin{equation}
 z_+^{(\varsigma)}\coloneqq -\mathsf X^{-1}y_+^{(\varsigma)},\qquad
 z_-^{(\varsigma)}\coloneqq \mathsf X^\dagger y_-^{(\varsigma)}.
 \label{eq:main-transformation}
\end{equation}
Substituting Eq.~\eqref{eq:main-transformation} into the chiral system
\eqref{eq:bg-chiral-system} and using Eq.~\eqref{eq:main-X}
and $\mathsf A^\dagger=\mathsf A$, we obtain
\begin{equation}
 \begin{aligned}
 \bigl(\mathscr D+(\alpha_r+\alpha_\theta)\sigma_2\bigr)z_+^{(\varsigma)}
     &=\mathsf C^{(\varsigma)}z_-^{(\varsigma)},\\
 \bigl(\overline{\mathscr D}-(\alpha_r+\alpha_\theta)\sigma_2\bigr)z_-^{(\varsigma)}
     &=\operatorname{adj}\mathsf C^{(\varsigma)}z_+^{(\varsigma)}.
 \end{aligned}
 \label{eq:global-transformed-system}
\end{equation}
In the second equation,
$\overline{\mathscr D}\mathsf X^\dagger
=\mathsf X^\dagger\mathsf A+(\alpha_r+\alpha_\theta)\sigma_2\mathsf X^\dagger$;
the unit determinant fixes the adjugate coupling.

Choose eigenvectors of $\sigma_2$,
\begin{equation}
 e_\pm\coloneqq \frac1{\sqrt2}\begin{pmatrix}1\\ \pm i\end{pmatrix},
 \qquad
 \sigma_2e_\pm=\pm e_\pm,\quad
 \sigma_1e_\pm=\pm i e_\mp,\quad
 \sigma_3e_\pm=e_\mp.
 \label{eq:sep-polarization-basis}
\end{equation}
The radial coupling and the term $(\alpha_r+\alpha_\theta)\sigma_2$ preserve these vectors,
whereas the angular coupling exchanges them. Rearrange the four
transformed components into two doublets,
\begin{equation}
 \begin{gathered}
 z_+^{(\varsigma)}=F_1^{(\varsigma)}e_-+G_2^{(\varsigma)}e_+,
 \qquad
 z_-^{(\varsigma)}=G_1^{(\varsigma)}e_-+F_2^{(\varsigma)}e_+,\\
 F^{(\varsigma)}\coloneqq
 \begin{pmatrix}F_1^{(\varsigma)}\\F_2^{(\varsigma)}\end{pmatrix},
 \qquad
 G^{(\varsigma)}\coloneqq
 \begin{pmatrix}G_1^{(\varsigma)}\\G_2^{(\varsigma)}\end{pmatrix}.
 \end{gathered}
 \label{eq:ang-doublets}
\end{equation}
This is a constant unitary change of components. Define the angular
operator
\begin{equation}
 \mathcal D_\theta^{(\varsigma)}
 \coloneqq ih\sigma_3\partial_\theta
 -C_1^{(\varsigma)}\sigma_1
 -C_3^{(\varsigma)}\sigma_2+\alpha_\theta\mathsf I.
 \label{eq:ang-operator}
\end{equation}
Then Eq.~\eqref{eq:global-transformed-system} becomes
\begin{equation}
 \begin{aligned}
 u\partial_r F^{(\varsigma)}
 &=\bigl(\mathcal D_\theta^{(\varsigma)}+\alpha_r\bigr)
       F^{(\varsigma)}
   +(C_0^{(\varsigma)}-C_2^{(\varsigma)})G^{(\varsigma)},\\
 u\partial_r G^{(\varsigma)}
 &=-\bigl(\mathcal D_\theta^{(\varsigma)}+\alpha_r\bigr)
       G^{(\varsigma)}
   +(C_0^{(\varsigma)}+C_2^{(\varsigma)})F^{(\varsigma)}.
 \end{aligned}
 \label{eq:ang-doublet-evolution}
\end{equation}
Both doublets involve the same angular operator; the remaining
coefficients depend only on $r$. We therefore seek separated
solutions of the form
\begin{equation}
 F^{(\varsigma)}(r,\theta)
   =R_1^{(\varsigma)}(r)S^{(\varsigma)}(\theta),
 \qquad
 G^{(\varsigma)}(r,\theta)
   =R_2^{(\varsigma)}(r)S^{(\varsigma)}(\theta).
 \label{eq:ang-product-ansatz}
\end{equation}
Here $R_1^{(\varsigma)}$ and $R_2^{(\varsigma)}$ are scalar radial
functions, and
$S^{(\varsigma)}(\theta)\coloneqq
(S_1^{(\varsigma)}(\theta),S_2^{(\varsigma)}(\theta))^{\mathsf T}$
is a common nonzero angular doublet, to be determined.
Substitution into Eq.~\eqref{eq:ang-doublet-evolution} gives
\begin{align}
 R_1^{(\varsigma)}\mathcal D_\theta^{(\varsigma)}S^{(\varsigma)}
 &=\Bigl(u\partial_r R_1^{(\varsigma)}
   -\alpha_rR_1^{(\varsigma)}
   -(C_0^{(\varsigma)}-C_2^{(\varsigma)})R_2^{(\varsigma)}
   \Bigr)S^{(\varsigma)},
 \label{eq:ang-product-first}\\
 R_2^{(\varsigma)}\mathcal D_\theta^{(\varsigma)}S^{(\varsigma)}
 &=\Bigl(-u\partial_r R_2^{(\varsigma)}
   -\alpha_rR_2^{(\varsigma)}
   +(C_0^{(\varsigma)}+C_2^{(\varsigma)})R_1^{(\varsigma)}
   \Bigr)S^{(\varsigma)}.
 \label{eq:ang-product-second}
\end{align}
For a nontrivial solution, at least one radial amplitude is
nonzero at some radius. The corresponding equation then requires
$\mathcal D_\theta^{(\varsigma)}S^{(\varsigma)}$ to be proportional
to $S^{(\varsigma)}$.
Since both depend only on $\theta$, the proportionality factor
is a separation constant $\lambda^{(\varsigma)}$, independent
of $r$ and $\theta$.
Thus the angular functions satisfy
$\mathcal D_\theta^{(\varsigma)}S^{(\varsigma)}
=\lambda^{(\varsigma)}S^{(\varsigma)}$,
or, in components,
\begin{equation}
  \begin{pmatrix}
  ih\partial_\theta+\alpha_\theta&
        -C_1^{(\varsigma)}+i C_3^{(\varsigma)}\\
  -C_1^{(\varsigma)}-i C_3^{(\varsigma)}&
        -ih\partial_\theta+\alpha_\theta
 \end{pmatrix}
 \begin{pmatrix}S_1^{(\varsigma)}\\S_2^{(\varsigma)}\end{pmatrix}
 =\lambda^{(\varsigma)}
 \begin{pmatrix}S_1^{(\varsigma)}\\S_2^{(\varsigma)}\end{pmatrix}.
 \label{eq:ang-two-component}
\end{equation}
The remaining radial relations give
\begin{equation}
  u\frac{d}{dr}\begin{pmatrix}R_1^{(\varsigma)}\\R_2^{(\varsigma)}\end{pmatrix}
 =\begin{pmatrix}
  \lambda^{(\varsigma)}+\alpha_r&C_0^{(\varsigma)}-C_2^{(\varsigma)}\\
  C_0^{(\varsigma)}+C_2^{(\varsigma)}&-\lambda^{(\varsigma)}-\alpha_r
 \end{pmatrix}
 \begin{pmatrix}R_1^{(\varsigma)}\\R_2^{(\varsigma)}\end{pmatrix}.
 \label{eq:ang-radial-system}
 \csname ltx@label\endcsname{eq:hor-radial-system}
\end{equation}
The corresponding $z_\pm^{(\varsigma)}$ are
\begin{equation}
 \begin{aligned}
 z_+^{(\varsigma)}(r,\theta)&=R_1^{(\varsigma)}(r)S_1^{(\varsigma)}(\theta)e_-+R_2^{(\varsigma)}(r)S_2^{(\varsigma)}(\theta)e_+,\\
 z_-^{(\varsigma)}(r,\theta)&=R_1^{(\varsigma)}(r)S_2^{(\varsigma)}(\theta)e_++R_2^{(\varsigma)}(r)S_1^{(\varsigma)}(\theta)e_-.
 \end{aligned}
 \label{eq:ang-separated-ansatz}
 \csname ltx@label\endcsname{eq:hor-doublet}
\end{equation}
Both transformed chiral equations follow from these two systems,
without dividing by an angular component.
For fixed $\omega,m,\varsigma$, we call the fields associated with
one angular eigenfunction $S^{(\varsigma)}$ and its two radial
amplitudes $R_1^{(\varsigma)},R_2^{(\varsigma)}$ a paired angular sector.

Thus an angular solution and a radial solution at the same
$\lambda^{(\varsigma)}$ give a solution of the transformed system.
Inverting Eq.~\eqref{eq:main-transformation} and then the combinations
in Eqs.~\eqref{eq:bg-chiral-plus} and \eqref{eq:bg-chiral-minus}
returns the radiative variables.
Separation means two coupled angular equations and two coupled radial
equations; the components within either system remain coupled.
This substitution is valid for complex $\omega$. We next construct $\mathsf X$, establish completeness for
real $\omega$, and determine the physical horizon conditions.

\subsection{Main result}
\label{sec:main-result}
\label{thm:main}

Under the assumptions of Eq.~\eqref{eq:bg-parameter-assumptions} and
the simple-root condition following it, we obtain the following results.
\begin{enumerate}
\item The matrix $\mathsf{X}(r,\theta)$ defined in Eq.~\eqref{eq:explicit-X}
is independent of $\omega,m,\varsigma$ and has the following properties:
\begin{itemize}
\item \textbf{Invertibility.}
$\det\mathsf X=1$.
\item \textbf{Transport equation.}
$\mathscr D\mathsf X=\mathsf A\mathsf X+
\mathsf X\bigl(\alpha_r(r)+\alpha_\theta(\theta)\bigr)\sigma_2$.
\item \textbf{Separation of the coupling.}
Its transformed coupling has the form \eqref{eq:main-separated-C}.
\item \textbf{Boundary regularity.}
Both $\mathsf{X}$ and $\mathsf{X}^{-1}$ are real analytic in square-root
boundary coordinates on
the exterior region including both horizons and the rotation axis.
\end{itemize}
These properties are verified in Sec.~\ref{sec:verification-transformation}.
Consequently, for either sign of spin weight and every complex $\omega$,
the ansatz \eqref{eq:ang-separated-ansatz} reduces the transformed
equations to the angular system \eqref{eq:ang-two-component} and
the radial system \eqref{eq:ang-radial-system}.
The explicit coefficients are given in
Secs.~\ref{sec:integrating-factor} and \ref{sec:source-normalization}.
\item For real $\omega$ and integer $m$, the angular operator in
Eq.~\eqref{eq:ang-two-component}, initially acting on smooth
compactly supported two-component functions in
$L^2((0,\pi),d\theta/h;\mathbb C^2)$, is essentially self-adjoint
and has compact resolvent.
Its complete eigenbasis yields, through
Eq.~\eqref{eq:ang-separated-ansatz}, a complete expansion of the
four-component transformed fields into paired sectors, each governed
by Eq.~\eqref{eq:ang-radial-system}.
This expansion is established in Sec.~\ref{sec:separated}.
\item The horizon regularity conditions of the original Newman--Penrose variables
preserve these paired sectors, with the exact boundary map derived
in Sec.~\ref{sec:horizons}.
\end{enumerate}
Here the square-root boundary coordinates are
$u=\sqrt{\Delta_r}$ at a horizon and $v=ah\sin\theta = a\sin\theta\sqrt{\Delta_{\theta}}$ at an angular pole.
Regularity of $\mathsf X$ in $u$ allows odd powers and is distinct
from physical smoothness in $r$; the latter requires the weights and
parity conditions of Sec.~\ref{sec:horizons}.
For complex $\omega$, the separation and boundary identities remain
valid, without an assertion of complete orthonormal angular modes.

The invertible transformations establish equivalence between each
radiative subsystem and the transformed system for arbitrary
complex $\omega$. For real $\omega$, angular completeness also
gives an equivalent expansion in separated modes. They do not
establish that every solution of the subsystem reconstructs a full
linearized Einstein--Maxwell perturbation, or that independently chosen
solutions for positive and negative spin weights describe the same
physical perturbation.

\subsection{Explicit form of the transformation}
\label{sec:explicit-transformation}
\label{app:matrix-coefficients}
\label{app:matrix-involution}

The matrix below satisfies the separation
conditions of Sec.~\ref{sec:target-transformation}.
We choose this representation so that the coefficient of $\omega$
in the radial coupling is proportional to $\mathsf I$, while
the mode-independent angular coupling is proportional to $i\sigma_3$.
The simpler matrix subsequently presented by Zhao~\cite{Zhao:2026separation}
is related to this representation by a frame transformation that
preserves separation.
Sec.~\ref{app:zhao} gives that matrix and its coefficient functions
in our conventions, together with the explicit frame transformation.
The analysis in this paper can also be carried out using
$\mathsf X_{\mathrm Z}$ in place of $\mathsf X$,
with the corresponding coefficient functions.

Our separation matrix is
\begin{equation}
 \mathsf X(r,\theta)\coloneqq
 \sqrt{\frac{1}{\mathcal N}}
 \begin{pmatrix}
  \beta_1&-i(\mathsf J_{11}\beta_1+\mathsf J_{12}\beta_2)\\
  \beta_2&-i(\mathsf J_{21}\beta_1+\mathsf J_{22}\beta_2)
 \end{pmatrix}
 =\sqrt{\frac{1}{\mathcal N}}
 \bigl(\boldsymbol\beta,\,-i\mathsf J\boldsymbol\beta\bigr).
 \label{eq:explicit-X}
\end{equation}
Here $\boldsymbol\beta\coloneqq(\beta_1,\beta_2)^{\mathsf T}$,
and the parentheses in the last expression denote the two columns.
The scalar quantities are defined by
\begin{align}
 \beta_1&\coloneqq b_0+b_3+b_{12},\qquad
 \beta_2\coloneqq b_0-b_3+\overline{b_{12}},
 \label{eq:explicit-X-columns}\\
 \mathcal N&\coloneqq
 \det\bigl(\boldsymbol\beta,-i\mathsf J\boldsymbol\beta\bigr)
 =-i\Bigl(
 \mathsf J_{21}\beta_1^2
  -2\mathsf J_{11}\beta_1\beta_2
  -\mathsf J_{12}\beta_2^2\Bigr).
 \label{eq:explicit-X-normalization}
\end{align}
We define
\begin{align}
 \nu&\coloneqq \sqrt{(Q^2\Xi-M^2)^2+4\ell^{-2}a^2M^2Q^2}>0,
 \label{eq:app-matrix-definitions}\\
 k_\theta&\coloneqq -1+\ell^{-2}a^2(1-2\cos^2\theta).
 \label{eq:app-matrix-ktheta}
\end{align}
We also use the positive scalar functions
\begin{equation}
 \mathcal S_r^2\coloneqq \nu^2(r^2+a^2)^2+4a^2M^2Q^2\Delta_r,
 \quad \mathcal S_r>0,\qquad
 H_\theta\coloneqq M^2v^2+a^2Q^2\Xi^2\cos^2\theta>0.
 \label{eq:short-positive-functions}
\end{equation}
The entries of the traceless matrix $\mathsf J$ are
\begin{align}
 \mathsf J_{11}
   &\coloneqq\frac{2Q\bigl((Mr-Q^2)v-iMau\cos\theta\bigr)}{\nu\Sigma},\qquad
 \mathsf J_{22}\coloneqq-\mathsf J_{11},
 \label{eq:app-matrix-J}\\
 \mathsf J_{12}
   &\coloneqq\frac{i}{\nu}\Biggl(
      \frac{Q^2(u+v)^2}{z^2}
      -\left(M-\frac{Q^2}{z}\right)^2+\ell^{-2}Q^2w^2
      \Biggr),
 \label{eq:app-matrix-J12}\\
 \mathsf J_{21}
   &\coloneqq-\frac{i}{\nu}\Biggl(
      \frac{Q^2(u-v)^2}{w^2}
      -\left(M-\frac{Q^2}{w}\right)^2+\ell^{-2}Q^2z^2
      \Biggr).
 \label{eq:app-matrix-J21}
\end{align}
The remaining scalar coefficients are
\begin{align}
 b_0&\coloneqq \frac{\nu(r^2+a^2)}{\mathcal S_r}
  +\frac{2a^2Q^2u^2}{\nu\mathcal S_r\Sigma}
       \bigl(M^2(1+\cos^2\theta)+Q^2\Xi\sin^2\theta\bigr)
  -\frac{2Qauv\cos\theta}{\Sigma\sqrt{H_\theta}},
 \label{eq:explicit-X-b0}\\
 b_3&\coloneqq \frac{2Q^2uv}{\nu\mathcal S_r\Sigma}
       \bigl(Q^2\Xi(r^2+a^2)+M^2(a^2-r^2)\bigr)
  +\frac{Qa\cos\theta}{\sqrt{H_\theta}}
       \left(k_\theta-\frac{2v^2}{\Sigma}\right),
 \label{eq:explicit-X-b3}\\
 b_{12}&\coloneqq \frac{2Qu}{\nu\mathcal S_r z^2}
  \Bigl(Ma\cos\theta\bigl(Q^2\Xi(r^2+a^2)+M^2(a^2-r^2)\bigr)\nonumber\\
 &\hspace{20mm}
       -ia^2(Mr-Q^2)\bigl(M^2(1+\cos^2\theta)+Q^2\Xi\sin^2\theta\bigr)\Bigr)
  +\frac{v}{\sqrt{H_\theta}}
       \left(M-\frac{2iQ^2a\cos\theta}{z^2}\right).
 \label{eq:explicit-X-b12}
\end{align}
Here $b_0,b_3$ are real, while $b_{12}$ is generally complex.
We retain $u=\sqrt{\Delta_r}$, $v=ah\sin\theta$,
$z=r+ia\cos\theta$, $w=r-ia\cos\theta$, and $\Sigma=zw$
from Sec.~\ref{sec:background}.
The assumptions $a>0$, $Q\ne0$, and $\ell>0$ ensure $\nu>0$.
Both $\mathcal S_r$ and $H_\theta$ remain positive on
the exterior region including both horizons and the rotation axis.
Equations~\eqref{eq:explicit-X}--\eqref{eq:explicit-X-b12}
specify the transformation entirely in terms of the background.

\subsection{Verification of the separation conditions}
\label{sec:verification-transformation}
\label{app:matrix}
We now verify the four properties listed in the first result of Sec.~\ref{sec:main-result}, as well as $\mathcal{N} \neq 0$.

\paragraph{\texorpdfstring{Auxiliary matrices $\mathsf E$ and $\mathsf W$.}{Auxiliary matrices E and W}}
To organize the calculations, introduce
the Hermitian matrix
\begin{equation}
 \mathsf E\coloneqq\frac{1}{\sqrt{H_\theta}}
 \begin{pmatrix}
 Qa\cos\theta\left(-k_\theta+\dfrac{2v(u+v)}{\Sigma}\right)
 &v\left(-M+\dfrac{2iQ^2a\cos\theta}{z^2}\right)\\[2mm]
 v\left(-M-\dfrac{2iQ^2a\cos\theta}{w^2}\right)
 &Qa\cos\theta\left(k_\theta+\dfrac{2v(u-v)}{\Sigma}\right)
 \end{pmatrix}.
 \label{eq:short-regular-angular-source}
 \csname ltx@label\endcsname{eq:direct-E2}
 \csname ltx@label\endcsname{eq:short-regular-angular-entries}
\end{equation}
Define the Hermitian matrix
\begin{equation}
 \begin{aligned}
 \mathsf W&\coloneqq
 \begin{pmatrix}
 b_0+b_3&b_{12}\\
 \overline{b_{12}}&b_0-b_3
 \end{pmatrix}
 +\mathsf E.
 \end{aligned}
 \label{eq:short-positive-metric}
 \csname ltx@label\endcsname{eq:short-regular-metric}
\end{equation}
Direct substitution gives the following algebraic identities:
\footnote{The following observation provides a guide to finding
$\mathsf X$.
One may first seek a traceless matrix $\mathsf J$ satisfying
the transport identity \eqref{eq:direct-J-transport},
together with Hermitian matrices $\mathsf W>0$ and $\mathsf E$
satisfying Eqs.~\eqref{eq:component-algebraic-identities}
and \eqref{eq:short-metric-identities}.
Equation~\eqref{eq:explicit-X} then gives a local candidate
where $\mathcal N\ne0$.
It satisfies Eqs.~\eqref{eq:direct-X-identities}--\eqref{eq:direct-X-E}
and $\det\mathsf X=1$, while
$\mathsf X^{-1}(\mathscr D\mathsf X-\mathsf A\mathsf X)$
is proportional to $\sigma_2$.
These conditions do not uniquely determine the candidate matrices.
Their remaining freedom can be explored heuristically to seek
a coefficient $\alpha_r(r)+\alpha_\theta(\theta)$ with real
$\alpha_r,\alpha_\theta$ and a coupling of the separated form
\eqref{eq:main-separated-C}.
Here $\mathsf W$ and $\mathsf E$ are chosen to implement
the radial and angular coupling conditions stated in
Sec.~\ref{sec:explicit-transformation}, respectively.
Global nonvanishing and boundary regularity require separate checks.}
\begin{align}
 \mathsf J^2&=\mathsf I,\qquad
 \det\mathsf W=1,\qquad
 \det\mathsf E=-1,
 \label{eq:component-algebraic-identities}\\
 \mathsf J\mathsf W\mathsf J^\dagger&=\mathsf W,\qquad
 \mathsf J\mathsf E\mathsf J^\dagger=-\mathsf E.
 \label{eq:short-metric-identities}
\end{align}

Since $\mathsf J$ is traceless, the relation
$\mathsf J^2=\mathsf I$ in
Eq.~\eqref{eq:component-algebraic-identities} also implies
\begin{equation}
 \det\mathsf J=-1.
 \label{eq:component-J-determinant}
\end{equation}
The relations above and the definition
\eqref{eq:explicit-X-columns} of $\boldsymbol\beta$ also imply%
\footnote{Indeed, let
$\mathsf B\coloneqq\mathsf W-\mathsf E$ and
$\bm e\coloneqq(1,1)^{\mathsf T}$.
Eqs.~\eqref{eq:component-algebraic-identities}
and \eqref{eq:short-metric-identities} give
$\det\mathsf B=0$ and therefore $\mathsf B$ is rank-one.
The rank-one identity
$(\mathsf B\bm e)(\mathsf B\bm e)^\dagger
=(\bm e^\dagger\mathsf B\bm e)\mathsf B$, with
$\boldsymbol\beta=\mathsf B\bm e$ and
$\bm e^\dagger\mathsf B\bm e
=2(b_0+\operatorname{Re}b_{12})$
then gives the result.}
\begin{equation}
 \boldsymbol\beta\boldsymbol\beta^\dagger
 =2\bigl(b_0+\operatorname{Re}b_{12}\bigr)
   (\mathsf W-\mathsf E).
 \label{eq:beta-outer-product}
\end{equation}

As shown below,
$|\mathcal N|=4\bigl(b_0+\operatorname{Re}b_{12}\bigr)>0$
throughout the exterior region including both horizons and the rotation axis, so Eq.~\eqref{eq:explicit-X}
defines $\mathsf X$ locally.
We next establish the relations between the auxiliary matrices
and $\mathsf X$:
\begin{align}
 \mathsf J\mathsf X&=\mathsf X\sigma_2,
 \label{eq:direct-X-identities}\\
 \mathsf W&=\mathsf X\mathsf X^\dagger,
 \label{eq:direct-X-W}\\
 \mathsf E&=-\mathsf X\sigma_3\mathsf X^\dagger.
 \label{eq:direct-X-E}
\end{align}
The first identity follows from
$\mathsf J^2=\mathsf I$ and the two-column form of $\mathsf X$
in Eq.~\eqref{eq:explicit-X}.
The remaining two follow from Eq.~\eqref{eq:beta-outer-product},
the conjugation identities in Eq.~\eqref{eq:short-metric-identities},
and the normalization
$|\mathcal N|=4\bigl(b_0+\operatorname{Re}b_{12}\bigr)$
established below.

\subsubsection{Invertibility}
\label{sec:determinant-transformation}
Using Eq.~\eqref{eq:beta-outer-product} and the conjugation
identities in Eq.~\eqref{eq:short-metric-identities}, we obtain
\begin{equation}
 \begin{aligned}
 &\bigl(\boldsymbol\beta,-i\mathsf J\boldsymbol\beta\bigr)
  \bigl(\boldsymbol\beta,-i\mathsf J\boldsymbol\beta\bigr)^\dagger
 =\boldsymbol\beta\boldsymbol\beta^\dagger
  +\mathsf J\boldsymbol\beta\boldsymbol\beta^\dagger\mathsf J^\dagger
 =4\bigl(b_0+\operatorname{Re}b_{12}\bigr)\mathsf W.
 \end{aligned}
 \label{eq:component-column-gram}
\end{equation}
The left-hand side is the product of the matrix whose columns
define $\mathcal N$ in Eq.~\eqref{eq:explicit-X-normalization}
and its Hermitian conjugate.
Taking determinants and using $\det\mathsf W=1$ gives
\begin{equation}
 |\mathcal N|^2=16\bigl(b_0+\operatorname{Re}b_{12}\bigr)^2.
 \label{eq:component-normalization-modulus}
\end{equation}
It therefore remains to show that
$b_0+\operatorname{Re}b_{12}>0$ throughout
the exterior region including both horizons and the rotation axis.
We first record two consequences of subextremality.
At the black hole outer horizon, $\Delta_r'(r_{\mathrm o})>0$ gives $r_{\mathrm o}>M$ and
\begin{equation}
 Mr_{\mathrm o}=a^2+Q^2+\ell^{-2}r_{\mathrm o}^4
       +\frac{r_{\mathrm o}\Delta_r'(r_{\mathrm o})}2>a^2+Q^2.
 \label{eq:component-horizon-inequality}
\end{equation}
The difference
\begin{equation}
 \Delta_r-(r-M)^2
   =a^2+Q^2-M^2-\ell^{-2}r^2(r^2+a^2)
 \label{eq:component-radial-inequality}
\end{equation}
is decreasing for $r>0$ and is negative at $r_{\mathrm o}$.
Consequently, throughout the exterior,
\begin{equation}
 0\le u<r-M,\qquad Mr>a^2+Q^2\ge2a|Q|.
 \label{eq:component-exterior-bounds}
\end{equation}
Set $\xi\coloneqq2Qa\cos\theta/\Sigma$. The radial bounds imply
$r|\xi|<M$ and
\begin{equation}
 M-u\xi-r\xi^2\ge(1+|\xi|)(M-r|\xi|)>0.
 \label{eq:component-seed-sign}
\end{equation}
The background definitions also give
\begin{equation}
 (r^2+a^2)^2-\left(\frac{uv}{h^2}\right)^2
 =\frac{\Xi(r^2+a^2)\Sigma
        +(2Mr-Q^2)a^2\sin^2\theta}{h^2}>0.
 \label{eq:component-column-weights}
\end{equation}
The right-hand side is positive by
Eq.~\eqref{eq:component-exterior-bounds}, and hence
$r^2+a^2\pm uv/h^2>0$.
A second direct substitution gives the manifestly positive expression
\begin{equation}
 \begin{aligned}
 b_0+\operatorname{Re}b_{12}
 ={}&\frac{\nu}{4\mathcal S_r}
 \Biggl(
 2(r^2+a^2)
 +\left(r^2+a^2+\frac{uv}{h^2}\right)
   |\mathsf J_{11}+\mathsf J_{21}|^2\\
 &\hspace{23mm}
 +\left(r^2+a^2-\frac{uv}{h^2}\right)
   |\mathsf J_{12}-\mathsf J_{11}|^2
 \Biggr)
 +\frac{v}{\sqrt{H_\theta}}(M-u\xi-r\xi^2)>0.
 \end{aligned}
 \label{eq:component-seed-positive}
\end{equation}
Here the first term in parentheses is strictly positive,
while all remaining terms are nonnegative by
Eqs.~\eqref{eq:component-seed-sign}
and \eqref{eq:component-column-weights}.
Combining this positivity with
Eq.~\eqref{eq:component-normalization-modulus} gives
\begin{equation}
 |\mathcal N|=4\bigl(b_0+\operatorname{Re}b_{12}\bigr)>0.
 \label{eq:component-nonzero-normalization}
\end{equation}
Thus Eq.~\eqref{eq:explicit-X} defines $\mathsf X$ locally
after either sign of the scalar square root is chosen at one
reference point and continued continuously.
The global analytic choice is established in
Sec.~\ref{sec:global-transformation}.

The determinant of the matrix multiplying the scalar factor in
Eq.~\eqref{eq:explicit-X} is $\mathcal N$. Hence
\begin{equation}
 \det\mathsf X=1.
 \label{eq:transformation-determinant}
\end{equation}
Thus $\mathsf X$ is invertible.
Equation~\eqref{eq:direct-X-W} then also implies
$\mathsf W>0$ throughout the exterior region including both horizons and the rotation axis.

\subsubsection{Transport equation}
\label{sec:integrating-factor}
\label{app:matrix-scalar-connection}
\label{app:matrix-trace-evaluation}

Differentiation of the entries in Eqs.~\eqref{eq:app-matrix-J}--\eqref{eq:app-matrix-J21} gives
\begin{equation}
 \mathscr D\mathsf J=[\mathsf A,\mathsf J].
 \label{eq:direct-J-transport}
\end{equation}
Differentiating $\mathsf J\mathsf X=\mathsf X\sigma_2$ now shows that
$\mathsf X^{-1}(\mathscr D\mathsf X-\mathsf A\mathsf X)$ commutes
with $\sigma_2$. Its trace vanishes because $\det\mathsf X$ is
constant and $\operatorname{tr}\mathsf A=0$.
It is therefore proportional to $\sigma_2$.

Denoting the proportionality coefficient by $\alpha$, we write
\begin{equation}
 \mathsf X^{-1}(\mathscr D\mathsf X-\mathsf A\mathsf X)
 =\alpha\sigma_2.
\end{equation}
This has the matrix structure required by
Eq.~\eqref{eq:main-X}; it remains to show that
$\alpha=\alpha_r(r)+\alpha_\theta(\theta)$
with real functions $\alpha_r$ and $\alpha_\theta$.
Using the first column of Eq.~\eqref{eq:explicit-X}
and $\det\mathsf X=1$, we obtain
\begin{equation}
 \begin{aligned}
 \alpha&=-\frac{i}{\mathcal N}
 \det\bigl(\boldsymbol\beta,
 \mathscr D\boldsymbol\beta-\mathsf A\boldsymbol\beta\bigr)\\
 &=-\frac{i}{\mathcal N}\Biggl(
 \beta_1\mathscr D\beta_2-\beta_2\mathscr D\beta_1
 -iQ\left(\dfrac{z}{w^2}\beta_1^2
          +\dfrac{w}{z^2}\beta_2^2\right)\Biggr).
 \end{aligned}
 \label{eq:direct-X-residual}
\end{equation}
The derivative of the common scalar factor cancels in this
combination. To evaluate the coefficient, define
\begin{equation}
 \mathcal Q_r\coloneqq \bigl(2Mr-\Xi(r^2+a^2)\bigr)
          \bigl((Q^2\Xi-M^2)r^2+a^2(M^2+Q^2\Xi)\bigr)
 +4a^2M^2(Mr-Q^2),
 \label{eq:short-radial-factorization}
\end{equation}
and set
\begin{equation}
 \alpha_r(r)\coloneqq \frac{Q\nu\mathcal Q_r}{2M\mathcal S_r^2}-\frac{Q\Xi}{2M},
 \qquad
 \alpha_\theta(\theta)\coloneqq \frac{Q\nu\Xi^2a^2\cos^2\theta}{2MH_\theta}+\frac{Q\Xi}{2M}.
 \label{eq:global-connection-functions}
\end{equation}
Substituting the explicit entries gives
\begin{equation}
 \alpha=\frac{Q\nu}{2M}
 \left(\frac{\mathcal Q_r}{\mathcal S_r^2}
       +\frac{\Xi^2a^2\cos^2\theta}{H_\theta}\right)=\alpha_r(r)+\alpha_\theta(\theta).
 \label{eq:direct-trace-value}
 \csname ltx@label\endcsname{eq:app-matrix-alpha-final}
\end{equation}
The real functions $\alpha_r,\alpha_\theta$ are regular on their closed
intervals; their constant terms fix the convention for
$\lambda^{(\varsigma)}$.
Thus the residual is $\bigl(\alpha_r(r)+\alpha_\theta(\theta)\bigr)\sigma_2$, as required in
Eq.~\eqref{eq:main-X}.
The final scalar denominators
$M,\nu,\mathcal S_r,H_\theta,\Sigma$ remain nonzero on
the exterior region including both horizons and the rotation axis;
regularity of $\mathsf X$ supplies the endpoint extension.

\subsubsection{Separation of the coupling}
\label{sec:source-normalization}
\label{app:matrix-algebraic-identities}

Expand $\mathsf C^{(\varsigma)}$ in the basis
$(\mathsf I,\sigma_2,i\sigma_1,i\sigma_3)$.
The identity $\operatorname{tr}(\sigma_j\sigma_k)=2\delta_{jk}$
and Eqs.~\eqref{eq:direct-X-identities}--\eqref{eq:direct-X-E} give its coefficients as
\begin{equation}
 \begin{aligned}
 C_0^{(\varsigma)}
   &= -\frac{1}{2}\operatorname{tr}
       (\mathsf W^{-1}\mathsf F^{(\varsigma)}),&
 C_2^{(\varsigma)}
   &= -\frac{1}{2}\operatorname{tr}
       (\mathsf W^{-1}\mathsf J\mathsf F^{(\varsigma)}),\\
 C_1^{(\varsigma)}
   &= \frac{1}{2}\operatorname{tr}
       (\mathsf E^{-1}\mathsf J\mathsf F^{(\varsigma)}),&
 C_3^{(\varsigma)}
   &= -\frac{i}{2}\operatorname{tr}
       (\mathsf E^{-1}\mathsf F^{(\varsigma)}).
 \end{aligned}
 \label{eq:direct-coefficient-extraction}
\end{equation}
Substituting the explicit matrices into the trace formulas gives
\begin{align}
 C_0^{(\varsigma)}
   &=\frac{3\nu\bigl(\frac{r^2+a^2}{2}\Delta_r'-2r\Delta_r\bigr)}{2u\mathcal S_r}
       \nonumber\\
   &\quad+\frac{i\varsigma\Xi}{\nu u\mathcal S_r}
       \Bigl(\omega\mathcal S_r^2
       +am\bigl(-\nu^2(r^2+a^2)+2Q^2(Q^2\Xi-M^2)\Delta_r\bigr)\Bigr),
 \label{eq:ang-coefficient-combinations}\\
 C_2^{(\varsigma)}
   &=-\frac{3}{2Q\mathcal S_r}
       \bigl(M^3(r^2+a^2)-2M^2Q^2r
       +MQ^2\Xi(a^2-r^2)+2Q^4\Xi r\bigr)
       \nonumber\\
   &\quad-\frac{2i\varsigma amQ\Xi(Mr-Q^2)}{\mathcal S_r},
 \label{eq:ang-coefficient-br}\\
 C_1^{(\varsigma)}
   &=\frac{\varsigma M\Xi}{\nu\sqrt{H_\theta}}
       \Bigl(-\omega a^2\bigl(M^2\sin^2\theta
       +Q^2\Xi(1+\cos^2\theta)\bigr)
       \nonumber\\
   &\hspace{30mm}
       +am\bigl(M^2-Q^2+\ell^{-2}a^2Q^2(1+2\cos^2\theta)\bigr)\Bigr),
 \label{eq:ang-coefficient-c}\\
 C_3^{(\varsigma)}
   &=\frac{3\sqrt{H_\theta}}{2Qv}
       +\frac{\varsigma Q\Xi a\cos\theta\bigl(\omega\Xi a^2\sin^2\theta-amk_\theta\bigr)}
                    {v\sqrt{H_\theta}}.
 \label{eq:ang-coefficient-d}
\end{align}
For fixed mode parameters, these expressions show that
$C_0^{(\varsigma)}$ and $C_2^{(\varsigma)}$ depend only on $r$,
whereas $C_1^{(\varsigma)}$ and
$C_3^{(\varsigma)}$ depend only on $\theta$.
Since the Pauli matrices and the identity span all complex
$2\times2$ matrices, these coefficients determine the coupling:
\begin{equation}
 \mathsf C^{(\varsigma)}
 =C_0^{(\varsigma)}\mathsf I+C_2^{(\varsigma)}\sigma_2
  +i C_1^{(\varsigma)}\sigma_1
  +i C_3^{(\varsigma)}\sigma_3.
 \label{eq:direct-coupling}
\end{equation}
This is the form required in Eq.~\eqref{eq:main-separated-C},
for complex $\omega$ and either spin label, since the
calculation is affine in $\omega,m$ and uses no conjugation of
these parameters. All coefficients are algebraic functions of the background and mode
parameters.

\subsubsection{Boundary regularity}
\label{sec:global-transformation}

The $1/u$ and $1/v$ factors are endpoint singularities of the
differential equations; however, the transformation $\mathsf X$
is regular there, as we now show.

The functions $\beta_1,\beta_2,\mathcal N$ and the entries of $\mathsf J$
are real analytic in the square-root boundary coordinates. Their denominators
$\nu,\mathcal S_r,\Sigma,z,w,\sqrt{H_\theta}$ are nonzero on
the exterior region including both horizons and the rotation axis, and Eq.~\eqref{eq:component-nonzero-normalization}
also gives $\mathcal N\ne0$ there. Since this domain is simply connected,
the scalar function $1/\mathcal N$ has an analytic square root fixed
by its sign at one point. Thus $\mathsf X$ extends analytically
to every boundary and corner, and its constant nonzero determinant
ensures that $\mathsf X^{-1}$ does so as well.

\subsection{Relation to Hintz's transformation}
\label{sec:local-kn-limit}

We now show how the separated formulation reduces locally
to Hintz's Kerr--Newman formulation as $\Lambda\to0$.
We compare the limiting transformation matrix with Hintz's
integrating factor and show that they are related by a purely
radial factor and a purely angular factor.
These factors preserve separation and establish the local
equivalence of the two formulations.
We then show that the separated residual can also be
removed at finite $\Lambda$ and explain why we retain it
in our algebraic formulation.

Fix $M>0$, $a>0$, and $Q\ne0$ with $M^2>a^2+Q^2$.
At fixed finite $r$ outside the Kerr--Newman black hole
outer horizon, the limit $\Lambda\to0$ gives
$\Delta_0\coloneqq r^2-2Mr+a^2+Q^2$ and
$\nu\to M^2-Q^2$.
Let $\mathsf X_0$ denote the corresponding limit of
Eq.~\eqref{eq:explicit-X}, with a consistent square-root branch.
We normalize Hintz's integrating factor in Eq.~(5) of
Ref.~\cite{Hintz:2026jiw} to unit determinant and write it as
\begin{equation}
 \mathsf X_{\mathrm{Hintz}}\coloneqq
 \frac{i}{\sqrt{M^2-Q^2}}
 \begin{pmatrix}
 \dfrac{Q(\sqrt{\Delta_0}+a\sin\theta)}{z}
 &-i\left(M-\dfrac{Q^2}{z}\right)\\
 i\left(M-\dfrac{Q^2}{w}\right)
 &\dfrac{Q(\sqrt{\Delta_0}-a\sin\theta)}{w}
 \end{pmatrix}.
 \label{eq:kn-hintz-matrix}
\end{equation}
Substituting $\Lambda=0$ into Eq.~\eqref{eq:explicit-X}
gives, with a matching choice of the overall square-root sign,
\begin{equation}
 \begin{aligned}
 \mathsf X_0
 &=\mathsf X_{\mathrm{Hintz}}e^{(f_0+ig_0)\sigma_2}=\mathsf X_{\mathrm{Hintz}}\bigl(
 \cosh(f_0+ig_0)\mathsf I+\sinh(f_0+ig_0)\sigma_2
 \bigr).
 \end{aligned}
 \label{eq:kn-matrix-relation}
\end{equation}
Here the real functions $f_0(r)$ and $g_0(\theta)$
are specified by
\begin{align}
 e^{2f_0(r)}
 &=\sqrt{
 \frac{\bigl(Mr-Q^2-Q\sqrt{\Delta_0}\bigr)^2+M^2a^2}
      {\bigl(Mr-Q^2+Q\sqrt{\Delta_0}\bigr)^2+M^2a^2}},
 \label{eq:kn-radial-factor}\\
 e^{2ig_0(\theta)}
 &=\frac{Q\cos\theta+iM\sin\theta}
        {\sqrt{M^2\sin^2\theta+Q^2\cos^2\theta}}.
 \label{eq:kn-angular-phase}
\end{align}
The square root in Eq.~\eqref{eq:kn-radial-factor} is positive,
and $g_0$ is chosen as a continuous half-argument of the phase
in Eq.~\eqref{eq:kn-angular-phase}.
Taking logarithmic derivatives of Eqs.~\eqref{eq:kn-radial-factor}
and \eqref{eq:kn-angular-phase} gives
\begin{equation}
 \frac{df_0}{dr}
 =\frac{\left.\alpha_r\right|_{\Lambda=0}}{\sqrt{\Delta_0}},
 \qquad
 \frac{dg_0}{d\theta}
 =\left.\alpha_\theta\right|_{\Lambda=0}
 =\frac{MQ}{2(M^2\sin^2\theta+Q^2\cos^2\theta)}.
 \label{eq:kn-factor-derivatives}
\end{equation}
The radial and angular factors preserve separation, so the
limiting formulation is locally equivalent to Hintz's.
This comparison does not address the limit of the cosmological
horizon boundary condition as that horizon recedes to infinity.

Equation~\eqref{eq:kn-factor-derivatives} also clarifies
the role of the separated residual in
Eq.~\eqref{eq:main-X}.
Hintz's transformation removes the matrix coefficient
$\mathsf A$ without leaving a residual, whereas ours leaves
$(\alpha_r+\alpha_\theta)\sigma_2$.
This residual can also be removed at finite $\Lambda$.
Define real functions $f(r),g(\theta)$ by
\begin{equation}
 \frac{df}{dr}=\frac{\alpha_r}{u},
 \qquad
 \frac{dg}{d\theta}=\frac{\alpha_\theta}{h}.
 \label{eq:finite-lambda-factor-derivatives}
\end{equation}
Then the matrix
\begin{equation}
 \widetilde{\mathsf X}\coloneqq
 \mathsf X e^{-(f+ig)\sigma_2}
 \label{eq:finite-lambda-integrating-factor}
\end{equation}
satisfies
\begin{equation}
 \begin{aligned}
 \mathscr D\widetilde{\mathsf X}
 &=\mathsf A\widetilde{\mathsf X}
 +\widetilde{\mathsf X}\bigl(
 (\alpha_r-u\partial_r f)
 +(\alpha_\theta-h\partial_\theta g)
 \bigr)\sigma_2\\
 &=\mathsf A\widetilde{\mathsf X},
 \end{aligned}
 \label{eq:finite-lambda-exact-transport}
\end{equation}
where the last equality follows from
Eq.~\eqref{eq:finite-lambda-factor-derivatives}.
To check the coupling, note that
$C_0^{(\varsigma)}\mathsf I+C_2^{(\varsigma)}\sigma_2$
commutes with $\sigma_2$, whereas
$iC_1^{(\varsigma)}\sigma_1+iC_3^{(\varsigma)}\sigma_3$
anticommutes with it. Therefore,
\begin{equation}
 \begin{aligned}
 \widetilde{\mathsf C}^{(\varsigma)}
 &\coloneqq -\widetilde{\mathsf X}^{-1}
 \mathsf F^{(\varsigma)}
 (\widetilde{\mathsf X}^{\dagger})^{-1}\\
 &=e^{2f(r)\sigma_2}
   \bigl(C_0^{(\varsigma)}(r)\mathsf I
        +C_2^{(\varsigma)}(r)\sigma_2\bigr)
    +e^{2ig(\theta)\sigma_2}
   \bigl(iC_1^{(\varsigma)}(\theta)\sigma_1
        +iC_3^{(\varsigma)}(\theta)\sigma_3\bigr).
 \end{aligned}
 \label{eq:finite-lambda-separated-coupling}
\end{equation}
The first term depends only on $r$ and remains a linear
combination of $\mathsf I,\sigma_2$.
The second depends only on $\theta$ and remains a linear
combination of $i\sigma_1,i\sigma_3$.
Thus the coupling retains the separated form
\eqref{eq:main-separated-C}.
These transformations illustrate the freedom in choosing
$\mathsf X$ while preserving the separated form of the equations.
We retain the separated residual because it can be incorporated
directly into the radial and angular equations.
This allows the transformation to remain explicitly algebraic,
without introducing the integrals defining $f$ and $g$.

The same freedom relates our transformation to the
simpler matrix of Zhao~\cite{Zhao:2026separation}.
In this case, elementary functions suffice to describe
the change of frame, while a separated residual remains.
The explicit relation is given in Sec.~\ref{app:zhao-relation}.

\section{Angular modes and completeness}
\label{sec:separated}
\label{sec:angular-expansion}

So far, we have constructed the invertible transformation
$\mathsf X$ and derived angular and radial equations whose
solutions combine to give separated solutions of the
transformed radiative system.
The angular functions satisfy
$\mathcal D_\theta^{(\varsigma)}S^{(\varsigma)}
=\lambda^{(\varsigma)}S^{(\varsigma)}$,
whose component form is given in Eq.~\eqref{eq:ang-two-component}.
We now establish that these separated solutions are sufficient
to represent the transformed radiative fields arising from
smooth physical perturbations: their general angular dependence
can be expressed as a superposition of angular eigenfunctions,
with each mode governed by an independent radial system.

We work at fixed real $\omega$, integer $m$, and spin label
$\varsigma$. To make this expansion precise, we consider
two-component angular functions whose squared magnitude
is integrable with weight $1/h$.
The corresponding function space and inner product are
\begin{equation}
 \mathcal H_{\mathrm{ang}}\coloneqq L^2((0,\pi),d\theta/h;\mathbb C^2),
 \qquad
 \langle F,G\rangle_{\mathrm{ang}}
 \coloneqq \int_0^\pi F^\dagger G\,\frac{d\theta}{h}.
 \label{eq:ang-inner-product}
\end{equation}
Here $L^2$ denotes square-integrable functions, and $\mathbb C^2$
indicates their two complex components.
Explicitly, $F^\dagger G=\overline{F_1}G_1+\overline{F_2}G_2$,
and a function belongs to $\mathcal H_{\mathrm{ang}}$ when
$\langle F,F\rangle_{\mathrm{ang}}<\infty$.
Since $1\le h(\theta)\le\sqrt\Xi$,
$\mathcal H_{\mathrm{ang}}$ consists of the same functions as
$L^2((0,\pi),d\theta;\mathbb C^2)$, with equivalent norms.
Our goal is to establish completeness of the angular eigenfunctions:
every function in $\mathcal H_{\mathrm{ang}}$ can be expanded
in eigenfunctions of $\mathcal D_\theta^{(\varsigma)}$,
with convergence in the norm defined by
Eq.~\eqref{eq:ang-inner-product}.

We use standard results from operator theory and verify
the conditions required to apply them to the angular operator
considered here.
We first analyze the behavior of angular solutions at the poles
to establish essential self-adjointness of the angular operator
(Sec.~\ref{sec:angular-domain}).
We then show that its resolvent is compact
(Sec.~\ref{sec:angular-spectrum}).
Together with self-adjointness, this guarantees a real
discrete spectrum and a complete orthonormal eigenbasis.
Finally, we apply this expansion to the radiative fields
(Sec.~\ref{sec:angular-field-expansion}).
For basic concepts and results in functional analysis,
including the definitions of dense subspaces, operator
adjoints, closures, self-adjointness, and compact
operators, we refer to Ref.~\cite{Teschl:2014}.
In particular, a symmetric operator is understood
to have a dense domain.

\subsection{Essential self-adjointness of the angular operator}
\label{sec:angular-domain}
\label{app:angular}

We first define the angular operator $\mathcal{D}_{\theta}^{(\varsigma)}$, given by Eq.~\eqref{eq:ang-operator}, on the initial domain
$C_c^\infty(0,\pi;\mathbb C^2)$, consisting of smooth
two-component functions that vanish in neighborhoods of both poles.
This domain is dense in $\mathcal H_{\mathrm{ang}}$
(see Theorem~0.40 of Ref.~\cite{Teschl:2014}).
Let $F$ and $G$ be arbitrary functions in this domain;
no differential equation is imposed on them.
In the inner product \eqref{eq:ang-inner-product},
the weight $1/h$ cancels the factor $h$ in
$ih\sigma_3\partial_\theta$.
For real $\omega$, the remaining coefficients are real
and the Pauli matrices are Hermitian.
Integration by parts, with vanishing boundary terms, therefore gives
$\langle F,\mathcal D_\theta^{(\varsigma)}G\rangle_{\mathrm{ang}}
=\langle\mathcal D_\theta^{(\varsigma)}F,G\rangle_{\mathrm{ang}}$.
Thus the operator is symmetric on this initial domain.

We use the kernel condition in
Corollary~2.8 of Ref.~\cite{Teschl:2014}.

\noindent\textbf{Criterion for essential self-adjointness.}
\textit{A symmetric operator $T$ is essentially
self-adjoint if and only if
the following condition holds for one
$\lambda\in\mathbb C\setminus\mathbb R$:}
\begin{equation}
 \operatorname{Ker}(T^*-\lambda)
 =\operatorname{Ker}(T^*-\bar\lambda)=\{0\}.
 \label{eq:ang-self-adjointness-criterion}
\end{equation}
Here $T^*$ denotes the adjoint of $T$
(see Eq.~(2.20) of Ref.~\cite{Teschl:2014}).
Eq.~\eqref{eq:ang-self-adjointness-criterion}
means that each of the equations
$(T^*-\lambda)S=0$ and $(T^*-\bar\lambda)S=0$
has only the zero solution $S=0$ in the domain of $T^*$.
Essential self-adjointness means that the closure
of $T$ is self-adjoint.
In this case, $T$ has a unique self-adjoint extension,
namely its closure, which coincides with $T^*$
(see Sec.~2.2 of Ref.~\cite{Teschl:2014},
in particular Lemma~2.4(ii)).

To apply this criterion to our angular system, we take
$T$ to be $\mathcal D_\theta^{(\varsigma)}$ initially defined
on $C_c^\infty(0,\pi;\mathbb C^2)$.
The adjoint has the same differential expression, but its
domain is larger.
Functions in the adjoint domain are not required to vanish
in neighborhoods of the poles.
Its maximal domain consists
of all functions $F\in\mathcal H_{\mathrm{ang}}$ for which
$\mathcal D_\theta^{(\varsigma)}F\in\mathcal H_{\mathrm{ang}}$,
with differentiation understood in the distributional sense.
Functions in this domain have locally absolutely continuous
representatives on $(0,\pi)$
(see Problem~2.31 in Sec.~2.7 of Ref.~\cite{Teschl:2014}).

We apply this criterion with $\lambda=i$.
It suffices to show that the angular equation has no
nonzero square-integrable solutions at $\lambda=\pm i$.
Local absolute continuity allows us to integrate by parts
on intervals away from the poles without multiplying the equation
by a compactly supported test function
(see Eq.~(2.119) of Ref.~\cite{Teschl:2014}).
We will exclude nonzero solutions by taking the interval
endpoints to the poles and showing that the boundary terms vanish.

We first identify the square-integrable
local solutions of the angular equation
with $\lambda=\pm i$ near each pole.
Let $\epsilon\coloneqq \theta$ at the north pole and
$\epsilon\coloneqq \pi-\theta$ at the south pole, and assign $\eta=+1,-1$
to these poles, respectively.
The background functions satisfy
\begin{equation}
 \cos\theta=\eta+O(\epsilon^2),\quad
 v=a\sqrt\Xi\,\epsilon+O(\epsilon^3),\quad
 H_\theta=Q^2\Xi^2a^2+O(\epsilon^2),\quad
 k_\theta=-\Xi+O(\epsilon^2).
\end{equation}
The coefficients in Eqs.~\eqref{eq:ang-coefficient-combinations}--\eqref{eq:ang-coefficient-d}
therefore give
\begin{equation}
 C_1^{(\varsigma)}=O(1),\qquad \alpha_\theta=O(1),\qquad
 C_3^{(\varsigma)}
 =\frac{\operatorname{sgn}(Q)\sqrt\Xi\,
             (\tfrac32+\eta\varsigma m)}{\epsilon}+O(\epsilon).
 \label{eq:ang-endpoint-coefficients}
\end{equation}
Using
$h=\sqrt\Xi+O(\epsilon^2)$, the angular equation becomes
\begin{equation}
 \frac{d S^{(\varsigma)}}{d\epsilon}
 =\left(-\frac{\operatorname{sgn}(Q)
                   (\varsigma m+\tfrac32\eta)}{\epsilon}\sigma_1
           +O(1)\right)S^{(\varsigma)}.
 \label{eq:ang-residue-matrix}
\end{equation}
The remainder is analytic at $\epsilon=0$. The two indicial exponents
are $\pm\beta_\eta^{(\varsigma)}$, where
\begin{equation}
 \beta_\eta^{(\varsigma)}\coloneqq |\varsigma m+\tfrac32\eta|\geq\tfrac12.
 \label{eq:ang-endpoint-exponents}
\end{equation}
Frobenius recursion gives one solution
$\epsilon^{\beta_\eta^{(\varsigma)}}(\boldsymbol v_0+O(\epsilon))$
with $\boldsymbol v_0\ne0$. The other has a nonzero
$\epsilon^{-\beta_\eta^{(\varsigma)}}$ leading term and may also
contain the first solution times $\log\epsilon$.
Only the first is square integrable, including the borderline case
$\beta_\eta^{(\varsigma)}=1/2$; the logarithmic term does not alter
this conclusion.

Any square-integrable solution with $\lambda=\pm i$
must therefore
select the square-integrable branch at both poles,
so its boundary current vanishes.
Integration by parts gives
\begin{equation}
 \pm 2i\,
       \|S^{(\varsigma)}\|_{\mathrm{ang}}^2
 =i\left[(S^{(\varsigma)})^\dagger
             \sigma_3S^{(\varsigma)}\right]_0^\pi=0.
 \label{eq:ang-deficiency-identity}
\end{equation}
Thus
$S^{(\varsigma)}=0$, so $T$ is essentially self-adjoint.
We henceforth use $\mathcal D_\theta^{(\varsigma)}$
to denote this self-adjoint closure.
Its domain is the maximal domain of the adjoint specified above.
No additional angular boundary condition is freely specifiable.

\subsection{Completeness of the angular eigenfunctions}
\label{sec:angular-spectrum}

We use the equivalence of conditions (i)--(iii) in
Proposition~5.12 of Ref.~\cite{Schmudgen:2012}.

\noindent\textbf{Self-adjoint operators with compact resolvent.}
\textit{Let $T$ be a self-adjoint operator on an
infinite-dimensional Hilbert space $\mathcal H$.
The following conditions are equivalent:}
\begin{enumerate}
 \item[(i)] \textit{
 $\mathcal H$ admits an orthonormal basis
 $\{e_n\}_{n\in\mathbb N}$ with
 $Te_n=\lambda_ne_n$, $\lambda_n\in\mathbb R$, and
 $|\lambda_n|\to\infty$ as $n\to\infty$.}
 \item[(ii)] \textit{
 The spectrum of $T$ is purely discrete.}
 \item[(iii)] \textit{
 $(T-z)^{-1}$ is compact at some $z\in\rho(T)$,
 and consequently at every $z\in\rho(T)$.}
\end{enumerate}
Here $\rho(T)$ denotes the resolvent set of $T$.
It consists of all $z\in\mathbb C$ for which
$(T-z)^{-1}$ exists as a bounded operator on all of
$\mathcal H$.
The spectrum of $T$ is defined as the complement
of $\rho(T)$ in $\mathbb C$.
Every eigenvalue belongs to the spectrum
(see Sec.~2.4 of Ref.~\cite{Teschl:2014}).

Self-adjointness of $\mathcal D_\theta^{(\varsigma)}$
was established in Sec.~\ref{sec:angular-domain}.
Since the spectrum of a self-adjoint operator is real
(see Theorem~2.19 of Ref.~\cite{Teschl:2014}),
every nonreal complex number belongs to the
resolvent set and, in particular, the resolvent
$(\mathcal D_\theta^{(\varsigma)}-i)^{-1}$ exists.
It maps a square-integrable source to the solution of the
inhomogeneous angular equation.
We construct this inverse from solutions of the
homogeneous angular equation and estimate its integral
kernel near both poles.
It therefore suffices to show that this resolvent is compact to prove completeness of the angular operator's eigenfunctions.

We first consider the homogeneous angular equation
\begin{equation}
 (\mathcal D_\theta^{(\varsigma)}-i)S^{(\varsigma)}=0.
 \label{eq:ang-resolvent-homogeneous}
\end{equation}
Let $S_N^{(\varsigma)}$ and
$S_S^{(\varsigma)}$ be nonzero solutions of this equation,
square integrable near
the north and south poles, respectively.
Extend both solutions to the full interval $(0,\pi)$.
These solutions are linearly independent: otherwise,
one nonzero solution would be square integrable at both poles,
contrary to the result of Sec.~\ref{sec:angular-domain}.
With $\lambda=i$, the equation in the $\theta$ coordinate is
\begin{equation}
 \partial_\theta S^{(\varsigma)}
 =\frac{C_1^{(\varsigma)}\sigma_2
        -C_3^{(\varsigma)}\sigma_1
        -i(\lambda-\alpha_\theta)\sigma_3}{h}S^{(\varsigma)}.
\end{equation}
Its coefficient matrix has zero trace. Thus the fundamental matrix
$\mathsf Y^{(\varsigma)}\coloneqq
 (S_N^{(\varsigma)},S_S^{(\varsigma)})$
has constant nonzero determinant.\footnote{
If the columns of $\mathsf Y^{(\varsigma)}$
were linearly dependent at one interior point, uniqueness
of the initial value problem would make the two solutions
linearly dependent throughout $(0,\pi)$, contradicting
their linear independence established above. Therefore $\det\mathsf Y^{(\varsigma)}\neq0$.
Its constancy follows from Jacobi's formula
$(\det\mathsf Y^{(\varsigma)})'
=\det\mathsf Y^{(\varsigma)}
\tr\bigl((\mathsf Y^{(\varsigma)})^{-1}
(\mathsf Y^{(\varsigma)})'\bigr)=0$,
where the last equality follows from the vanishing trace
of the coefficient matrix.
}

For a compactly supported two-component source
$f\in\mathcal H_{\mathrm{ang}}$, we now consider
the inhomogeneous equation
\begin{equation}
 (\mathcal D_\theta^{(\varsigma)}-i)S^{(\varsigma)}=f.
 \label{eq:ang-resolvent-inhomogeneous}
\end{equation}
We apply variation
of constants and choose the coefficient of $S_N^{(\varsigma)}$
to vanish near the south pole and that of $S_S^{(\varsigma)}$
to vanish near the north pole.
Integrating these coefficients from the respective poles
gives the
resolvent kernel relative to $d\theta'/h(\theta')$:
\begin{equation}
 \mathsf K^{(\varsigma)}(\theta,\theta')=
 \begin{cases}
 i S_N^{(\varsigma)}(\theta)
       (1,0)\mathsf Y^{(\varsigma)}(\theta')^{-1}\sigma_3,
       &\theta<\theta',\\
 -i S_S^{(\varsigma)}(\theta)
       (0,1)\mathsf Y^{(\varsigma)}(\theta')^{-1}\sigma_3,
       &\theta>\theta'.
 \end{cases}
 \label{eq:ang-green-kernel}
\end{equation}
Writing $S_{N,j}^{(\varsigma)}$ and
$S_{S,j}^{(\varsigma)}$ for the components of the two solutions,
we can equivalently express the kernel as
\begin{equation}
 \mathsf K^{(\varsigma)}(\theta,\theta')
 =\frac{i}{\det\mathsf Y^{(\varsigma)}}
 \begin{cases}
 \begin{pmatrix}
 S_{N,1}^{(\varsigma)}(\theta)\\
 S_{N,2}^{(\varsigma)}(\theta)
 \end{pmatrix}
 \begin{pmatrix}
 S_{S,2}^{(\varsigma)}(\theta')&
 S_{S,1}^{(\varsigma)}(\theta')
 \end{pmatrix},
 &\theta<\theta',\\[3mm]
 \begin{pmatrix}
 S_{S,1}^{(\varsigma)}(\theta)\\
 S_{S,2}^{(\varsigma)}(\theta)
 \end{pmatrix}
 \begin{pmatrix}
 S_{N,2}^{(\varsigma)}(\theta')&
 S_{N,1}^{(\varsigma)}(\theta')
 \end{pmatrix},
 &\theta>\theta'.
 \end{cases}
 \label{eq:ang-green-kernel-components}
\end{equation}
Define
\begin{equation}
 S^{(\varsigma)}(\theta)\coloneqq
 \int_0^\pi \mathsf K^{(\varsigma)}(\theta,\theta')
 f(\theta')\,\frac{d\theta'}{h(\theta')}.
 \label{eq:ang-resolvent-action}
\end{equation}
This function is square integrable at both poles and
satisfies Eq.~\eqref{eq:ang-resolvent-inhomogeneous}.
It therefore equals the resolvent applied to $f$.

We now show that the kernel is square integrable
in its two variables, meaning that
\begin{equation}
 \int_0^\pi\!\!\int_0^\pi
 \sum_{j,k=1}^{2}
 \left|\mathsf K_{jk}^{(\varsigma)}(\theta,\theta')\right|^2
 \frac{d\theta}{h(\theta)}
 \frac{d\theta'}{h(\theta')}<\infty.
 \label{eq:ang-kernel-square-integrability}
\end{equation}
For $\theta<\theta'$, Eq.~\eqref{eq:ang-green-kernel-components}
bounds each component in absolute value by
$C|S_N^{(\varsigma)}(\theta)|
   |S_S^{(\varsigma)}(\theta')|$.
Near the north pole this is at most
$C(\theta/\theta')^{\beta_+^{(\varsigma)}}\leq C$; near the south
pole it is at most
$C\bigl((\pi-\theta')/(\pi-\theta)\bigr)^{\beta_-^{(\varsigma)}}\leq C$.
The opposite triangle has the same bounds. Away from these corners
the solutions are bounded, or their endpoint-regular factors decay.
Thus the kernel is bounded for $\theta\neq\theta'$.
Its values on the diagonal do not affect the double integral,
since the diagonal has measure zero.
Since $\int_0^\pi d\theta/h<\infty$, it is square integrable.

An integral operator with a square-integrable kernel
is compact
(see Sec.~6.3 of Ref.~\cite{Teschl:2014}).
So far, the integral representation has been established
only for compactly supported sources $f$, but it must hold
for every square-integrable source
$f\in\mathcal H_{\mathrm{ang}}$.
Since the integral operator and the resolvent are bounded,
their agreement on compactly supported sources extends
by density to all of $\mathcal H_{\mathrm{ang}}$
(see Theorem~0.29 of Ref.~\cite{Teschl:2014}).
The resolvent is therefore compact, giving a real discrete
spectrum and a complete orthonormal eigenbasis
in $\mathcal H_{\mathrm{ang}}$.

\subsection{Expansion of the radiative fields}
\label{sec:angular-field-expansion}

We now apply the complete angular eigenbasis to the
doublets $F^{(\varsigma)}$ and $G^{(\varsigma)}$ in
Eq.~\eqref{eq:ang-doublets} obtained from smooth physical
perturbations.
After verifying that the doublets are square integrable
and lie in the domain of the angular operator, we expand
them in angular eigenfunctions and project their field
equations to obtain the radial equations.

For a smooth physical perturbation, a spin-weight $s$ field with
azimuthal number $m$ is $O(\theta^{|m+s|})$ at the north axis and
$O((\pi-\theta)^{|m-s|})$ at the south axis. The original radiative
fields have spin weights
$(2,1,1,2)$ for $\varsigma=+1$ and
$(-2,-1,-1,-2)$ for $\varsigma=-1$. Their chiral factor
$\sqrt{h\sin\theta}$ and the bounded matrices $\mathsf X$
and $\mathsf X^{-1}$ (see Sec.~\ref{sec:global-transformation})
therefore place both doublets in $\mathcal H_{\mathrm{ang}}$.
Their radial derivatives belong to the same space, uniformly on compact
radial subintervals.\footnote{
These uniform bounds, together with the assumed smoothness,
justify interchanging radial differentiation with the angular
inner products defining the expansion coefficients below.
}
Equation~\eqref{eq:ang-doublet-evolution} then
puts $\mathcal D_\theta^{(\varsigma)}F^{(\varsigma)}$
and $\mathcal D_\theta^{(\varsigma)}G^{(\varsigma)}$
in that space as well. Thus the doublets lie in the maximal
self-adjoint domain.

Choose a complete orthonormal eigenbasis
$\{S_j^{(\varsigma)}\}$, with real eigenvalues
$\lambda_j^{(\varsigma)}$ and with multiplicities included.
By Theorem~1.3 of Ref.~\cite{Teschl:2014}, we have
\begin{equation}
 F^{(\varsigma)}(r,\theta)
 =\sum_j R_{1j}^{(\varsigma)}(r)S_j^{(\varsigma)}(\theta),
 \qquad
 G^{(\varsigma)}(r,\theta)
 =\sum_j R_{2j}^{(\varsigma)}(r)S_j^{(\varsigma)}(\theta).
 \label{eq:ang-mode-expansion}
\end{equation}
Projection of Eq.~\eqref{eq:ang-doublet-evolution} onto each
eigenfunction gives Eq.~\eqref{eq:ang-radial-system} with
$\lambda^{(\varsigma)}=\lambda_j^{(\varsigma)}$.
Inserting Eq.~\eqref{eq:ang-mode-expansion} into
Eq.~\eqref{eq:ang-doublets} gives a sum of the paired products in
Eq.~\eqref{eq:ang-separated-ansatz}. This proves completeness of
that ansatz.

Zero or repeated eigenvalues are already covered by the preceding argument.
This implication starts from smooth physical fields. Conversely,
we have not established that arbitrary separated data yield radiative
fields smooth at both angular poles. The completeness statement here
is restricted to real $\omega$.

\section{Radiative fields and horizon boundary conditions}
\label{sec:horizons}

We now express smoothness of the radiative fields at the
future black hole outer and cosmological horizons as boundary
conditions on the separated radial amplitudes.
These future-horizon regularity conditions correspond
to the ingoing condition at the black hole outer horizon
and the outgoing condition at the cosmological horizon
used to define quasinormal modes.
The opposite choices---outgoing at the black hole outer
horizon and ingoing at the cosmological horizon---can be
studied by imposing regularity at the respective past
horizons.
The resulting conditions are given by
Eqs.~\eqref{eq:hor-common-weight} and
\eqref{eq:hor-radial-parity} and preserve each separated
angular sector.

We derive these conditions in three steps.
Sec.~\ref{sec:radiative-identification} expresses smoothness
in regular coordinates and tetrads as conditions on
$y_\pm^{(\varsigma)}$.
Sec.~\ref{sec:horizon-reflection} transfers these conditions
to $z_\pm^{(\varsigma)}$ using the transformation matrix
$\mathsf X$.
Sec.~\ref{sec:radial-boundary-conditions} then substitutes
the separated ansatz to obtain the radial boundary conditions.

\subsection{Radiative fields and horizon weights}
\label{sec:radiative-identification}

Our first task is to express horizon regularity in terms
of $y_\pm^{(\varsigma)}$.
We identify coordinates and tetrads regular at each future
horizon, require the radiative components in those frames
to be smooth, and substitute their expressions into the
definitions of $y_\pm^{(\varsigma)}$.
This gives the horizon forms in
Eq.~\eqref{eq:hor-chiral-table}.

Let $r_{\mathrm H}$ be either simple horizon, $r_{\mathrm o}$ or $r_{\mathrm c}$, approached from
$\Delta_r>0$.
The subscript $\mathrm H$ denotes evaluation at this horizon.
With Fourier convention $e^{-i\omega t+im\phi}$, define
\begin{equation}
 \frac{dr_*}{dr}=\frac{\Xi(r^2+a^2)}{\Delta_r},\qquad
 \Omega_{\mathrm H}\coloneqq \frac{a}{r_{\mathrm H}^2+a^2},\qquad k_{\mathrm H}\coloneqq \omega-m\Omega_{\mathrm H}.
 \label{eq:hor-frequency}
\end{equation}
Here $r_*\to-\infty$ at $r_{\mathrm o}$ and $r_*\to+\infty$ at $r_{\mathrm c}$.
The exact ingoing ($\eta=-1$) and outgoing ($\eta=+1$) phases are
\begin{equation}
 \mathcal P_\eta(r)
 \coloneqq \exp\!\left(-\eta i\int^r\frac{K(s)}{\Delta_r(s)}\,ds\right),
 \qquad K=\Xi\bigl(-(r^2+a^2)\omega+am\bigr).
 \label{eq:hor-phase}
\end{equation}
They equal $e^{\eta i k_{\mathrm H} r_*}$ up to smooth nonzero factors.
The coordinates regular at the future black hole outer
and cosmological horizons are, respectively,
\begin{align}
 t_{\mathrm{o}}&\coloneqq t+r_*,
 &\phi_{\mathrm{o}}&\coloneqq
 \phi+\int^r\frac{\Xi a}{\Delta_r(s)}\,ds,
 \label{eq:hor-coordinates-outer}\\
 t_{\mathrm{c}}&\coloneqq t-r_*,
 &\phi_{\mathrm{c}}&\coloneqq
 \phi-\int^r\frac{\Xi a}{\Delta_r(s)}\,ds.
 \label{eq:hor-coordinates-cosmological}
\end{align}
The phase factors satisfy
\begin{align}
 e^{-i\omega t+im\phi}\mathcal P_-
 &=e^{-i\omega t_{\mathrm{o}}+im\phi_{\mathrm{o}}},
 \label{eq:hor-phase-outer}\\
 e^{-i\omega t+im\phi}\mathcal P_+
 &=e^{-i\omega t_{\mathrm{c}}+im\phi_{\mathrm{c}}}.
 \label{eq:hor-phase-cosmological}
\end{align}
The future black hole outer and cosmological horizon conditions use $\eta=-1$
and $\eta=+1$, respectively, at the physical $\omega,m$ for either
spin label.

A component has boost weight $b$ if it
scales by $B^b$ under $\bm{k}_{\mathrm{NP}}\mapsto B \bm{k}_{\mathrm{NP}}$,
$\bm{l}_{\mathrm{NP}}\mapsto B^{-1}\bm{l}_{\mathrm{NP}}$.
Write a component of boost weight $b$ in the original tetrad
as $e^{-i\omega t+im\phi}f(r,\theta)$.

First consider the future black hole outer horizon at $r=r_{\mathrm o}$.
In the ingoing coordinates $(t_{\mathrm{o}},r,\theta,\phi_{\mathrm{o}})$,
the regular null vectors are
\begin{equation}
 \Delta_r \bm{k}_{\mathrm{NP}}=2\Xi(r^2+a^2)\bm{\partial}_{t_{\mathrm{o}}}
 +\Delta_r\bm{\partial}_r+2\Xi a\bm{\partial}_{\phi_{\mathrm{o}}},\qquad
 \Delta_r^{-1}\bm{l}_{\mathrm{NP}}=-\frac1{2\Sigma}\bm{\partial}_r.
\end{equation}
The Fourier amplitude $f_{\mathrm{o}}$ in these regular coordinates
and tetrad is defined by
\begin{equation}
 \Delta_r^b e^{-i\omega t+im\phi}f(r,\theta)
 =e^{-i\omega t_{\mathrm{o}}+im\phi_{\mathrm{o}}}f_{\mathrm{o}}(r,\theta).
 \label{eq:hor-regular-amplitudes}
\end{equation}
Regularity of the perturbation at this horizon requires
$f_{\mathrm{o}}$ to extend smoothly through $r=r_{\mathrm o}$.

Next consider the future cosmological horizon at $r=r_{\mathrm c}$.
Expressing Eqs.~\eqref{eq:bg-tetrad-k} and
\eqref{eq:bg-tetrad-l} in the outgoing coordinates
$(t_{\mathrm{c}},r,\theta,\phi_{\mathrm{c}})$ gives
\begin{align}
 \bm{k}_{\mathrm{NP}}&=\bm{\partial}_r,
 \label{eq:hor-outgoing-tetrad-k}\\
 \bm{l}_{\mathrm{NP}}&=
 \frac{2\Xi(r^2+a^2)\bm{\partial}_{t_{\mathrm{c}}}
       -\Delta_r\bm{\partial}_r
       +2\Xi a\bm{\partial}_{\phi_{\mathrm{c}}}}{2\Sigma}.
 \label{eq:hor-outgoing-tetrad-l}
\end{align}
The angular null vector in Eq.~\eqref{eq:bg-tetrad}
retains its form with $t,\phi$ replaced by $t_{\mathrm{c}},\phi_{\mathrm{c}}$.
The original tetrad therefore extends regularly through
$r=r_{\mathrm c}$ without a boost.
In these coordinates, write the same component as
$e^{-i\omega t_{\mathrm{c}}+im\phi_{\mathrm{c}}}f_{\mathrm{c}}(r,\theta)$.
Regularity at this horizon requires $f_{\mathrm{c}}$ to extend smoothly
through $r=r_{\mathrm c}$.

Using the phase relation above, the original Fourier
amplitude is therefore related to these regular amplitudes by
\begin{equation}
 f(r,\theta)=
 \begin{cases}
  \Delta_r^{-b}\mathcal P_-(r)f_{\mathrm{o}}(r,\theta),
    & \text{near }r=r_{\mathrm o},\\
  \mathcal P_+(r)f_{\mathrm{c}}(r,\theta),
    & \text{near }r=r_{\mathrm c}.
 \end{cases}
 \label{eq:hor-mode-amplitudes}
\end{equation}
The normalized variables have weights
\begin{equation}
\begin{array}{c|cccc}
\text{variables}
 &(\widehat{\Psi}_0,\widehat{\kappa})&(\widehat{\Psi}_1,\widehat{\sigma})&(\widehat{\Psi}_4,\widehat{\nu})&(\widehat{\Psi}_3,\widehat{\lambda})\\ \hline
b&2&1&-2&-1 .
\end{array}
\label{eq:hor-boost-weights}
\end{equation}
Their factors involving $r\pm ia\cos\theta$ are smooth and nonzero at the
horizons. The perturbed spin coefficients vanish on the background
and transform homogeneously with the displayed boost weights.

At fixed $0<\theta<\pi$, substituting the component
expressions in Eq.~\eqref{eq:hor-mode-amplitudes},
with the boost weights in Eq.~\eqref{eq:hor-boost-weights},
into the definitions of $y_\pm^{(\varsigma)}$ in
Eqs.~\eqref{eq:bg-chiral-plus} and \eqref{eq:bg-chiral-minus}
gives
\begin{equation}
\begin{array}{c|c|c}
 \varsigma&r_{\mathrm H}&y_\pm^{(\varsigma)}\\ \hline
 +1&r_{\mathrm o}&
 \mathcal P_-\Delta_r^{-3/4}
 (\mathbf a\pm i\sqrt{\Delta_r}\mathbf b)\\
 +1&r_{\mathrm c}&
 \mathcal P_+\Delta_r^{3/4}
 (\sqrt{\Delta_r}\mathbf a\pm i\mathbf b)\\
 -1&r_{\mathrm o}&
 \mathcal P_-\Delta_r^{3/4}
 (\sqrt{\Delta_r}\mathbf a\mp i\mathbf b)\\
 -1&r_{\mathrm c}&
 \mathcal P_+\Delta_r^{-3/4}
 (\mathbf a\mp i\sqrt{\Delta_r}\mathbf b).
\end{array}
\label{eq:hor-chiral-table}
\csname ltx@label\endcsname{eq:hor-physical-parity}
\end{equation}
In each row,
$\mathbf a,\mathbf b$ are two-component amplitudes
smooth in $r$ through the relevant horizon.
They are defined separately in each row but are common
to $y_+^{(\varsigma)}$ and $y_-^{(\varsigma)}$ within that row.
For the reflection argument below, we write
$\sqrt{\Delta_r}=u$.
Since $r$ is analytic in $u^2$ near a simple horizon,
these amplitudes are even in $u$.
Under $u\mapsto-u$, the parenthesized expressions for
$y_+^{(\varsigma)}$ and $y_-^{(\varsigma)}$ are therefore
exchanged with a factor $-\eta\varsigma$.
Negative $u$ encodes this parity and supplies
no additional physical region or boundary data.

\subsection{Reflection identities and physical regularity}
\label{sec:horizon-reflection}

We next express the horizon forms in
Eq.~\eqref{eq:hor-chiral-table} in terms of
$z_\pm^{(\varsigma)}$.
The key step is to determine how $\mathsf X$ transforms
under $u\mapsto-u$.
This yields the regularity condition
\eqref{eq:hor-master-parity}; the inverse transformation
then shows that it is equivalent to the original
horizon forms.

Near a simple horizon, $r=r(u^2)$ is analytic. Reflection
$u\mapsto-u$ therefore keeps $r$ and $\theta$ fixed.
The entries in Eqs.~\eqref{eq:app-matrix-J}--\eqref{eq:app-matrix-J21},
\eqref{eq:short-positive-metric}, and
\eqref{eq:short-regular-angular-source} give
\begin{equation}
 \mathsf J(-u)=\mathsf J(u)^\dagger,\qquad
 \mathsf E(-u)=-\operatorname{adj}\mathsf E(u),
\end{equation}
and
\begin{equation}
 \mathsf W(-u)=\operatorname{adj}\mathsf W(u)
             =\mathsf W(u)^{-1}.
\end{equation}
Consequently, $(\mathsf X(u)^\dagger)^{-1}$ has determinant
$1$ and satisfies the three identities
\eqref{eq:direct-X-identities}--\eqref{eq:direct-X-E} at $-u$.
These identities determine $\mathsf X$ up to sign: any two
solutions differ on the right by a unitary matrix commuting
with both $\sigma_2$ and $\sigma_3$, and their equal determinants
restrict this matrix to $\pm\mathsf I$.

At either horizon, setting $u=0$ in
Eq.~\eqref{eq:short-positive-metric} gives
$\mathsf W_{\mathrm H}=\mathsf I$.
Together with Eq.~\eqref{eq:direct-X-W}, this yields
$(\mathsf X_{\mathrm H}^\dagger)^{-1}=\mathsf X_{\mathrm H}$,
fixing the positive sign. Continuity and
$\det\mathsf X=1$ now give
\begin{equation}
 \mathsf X(-u)=\operatorname{adj}\mathsf X(u)^\dagger,
 \qquad
 \operatorname{adj}\mathsf X(-u)=\mathsf X(u)^\dagger.
 \label{eq:hor-X-parity}
\end{equation}
The reflection identities in Eq.~\eqref{eq:hor-X-parity}
remain valid at the angular poles because
$\mathsf E$ is regular there.

Applying the transformation \eqref{eq:main-transformation}
to the horizon forms in Eq.~\eqref{eq:hor-chiral-table} gives
\begin{equation}
 z_\pm^{(\varsigma)}
 =\Delta_r^{3\eta\varsigma/4}\mathcal P_\eta
 \widehat z_\pm^{(\varsigma)},
 \label{eq:hor-z-amplitudes}
\end{equation}
with amplitudes $\widehat z_\pm^{(\varsigma)}$ smooth in $u$.
The exchange property of the parenthesized expressions
in Eq.~\eqref{eq:hor-chiral-table},
together with Eq.~\eqref{eq:hor-X-parity}, implies
\begin{equation}
 \widehat z_-^{(\varsigma)}(u)
 =\eta\varsigma\widehat z_+^{(\varsigma)}(-u).
 \label{eq:hor-master-parity}
\end{equation}
Equivalently,
\begin{equation}
 \widehat z_+^{(\varsigma)}+\eta\varsigma\widehat z_-^{(\varsigma)},
 \qquad
 \frac{\widehat z_+^{(\varsigma)}-\eta\varsigma\widehat z_-^{(\varsigma)}}{u}
 \quad\text{are smooth even functions of }u.
 \label{eq:hor-even-conditions}
\end{equation}
These conditions hold at every order. Both $\operatorname{adj}\mathsf X$
and $\mathsf X^\dagger$ are smooth and invertible at $u=0$;
Eq.~\eqref{eq:hor-X-parity} also maps the parity conditions back to
the horizon forms in Eq.~\eqref{eq:hor-chiral-table}. Thus, at fixed $0<\theta<\pi$, the weighted transformation
preserves horizon regularity in both directions.

The inverse map from the separated amplitudes to the radiative fields is
\begin{equation}
 y_+^{(\varsigma)}=-\mathsf{X}z_+^{(\varsigma)},\qquad y_-^{(\varsigma)}=(\mathsf{X}^\dagger)^{-1}z_-^{(\varsigma)} ,
 \label{eq:discussion-inverse-chiral}
\end{equation}
followed by the inverse of Eqs.~\eqref{eq:bg-chiral-plus} and
\eqref{eq:bg-chiral-minus}. In particular,
\begin{equation}
 \Psi_4^{(1)}
 =\frac{\Delta_r^{3/4}}{2w^4\sqrt{h\sin\theta}}
 \left[-\mathsf{X}z_+^{(-)}+(\mathsf{X}^\dagger)^{-1}z_-^{(-)}\right]_1,
 \label{eq:discussion-Psi4}
\end{equation}
where the subscript $1$ selects the first component.

\subsection{Boundary conditions for separated radial amplitudes}
\label{sec:radial-boundary-conditions}

We finally substitute the separated ansatz into the
regularity condition for $z_\pm^{(\varsigma)}$.
We show that it reduces to a relation between the two
radial amplitudes in each separated angular sector,
without coupling different sectors.

For the separated radial functions
$R_j^{(\varsigma)}$ with $j=1,2$, write
\begin{equation}
 R_j^{(\varsigma)}
 =\Delta_r^{3\eta\varsigma/4}\mathcal P_\eta
 \widehat R_j^{(\varsigma)}.
 \label{eq:hor-common-weight}
\end{equation}
Substituting Eq.~\eqref{eq:ang-separated-ansatz} into
Eq.~\eqref{eq:hor-master-parity} gives
\begin{equation}
 \begin{aligned}
 \bigl(\widehat R_2^{(\varsigma)}(u)-\eta\varsigma
  \widehat R_1^{(\varsigma)}(-u)\bigr)S_1^{(\varsigma)}&=0,\\
 \bigl(\widehat R_1^{(\varsigma)}(u)-\eta\varsigma
  \widehat R_2^{(\varsigma)}(-u)\bigr)S_2^{(\varsigma)}&=0.
 \end{aligned}
\end{equation}
A nonzero angular spinor has at least one component that is not
identically zero. These equations and their reflections are therefore
equivalent to
\begin{equation}
 \widehat R_2^{(\varsigma)}(u)
 =\eta\varsigma\widehat R_1^{(\varsigma)}(-u),
 \label{eq:hor-radial-parity}
\end{equation}
with smooth hatted amplitudes in $u$. No division by an angular
component is needed, and each separated pair is preserved, including
at a zero or degenerate angular eigenvalue. For real $\omega$, the
same statement follows term by term in the complete expansion in
Eq.~\eqref{eq:ang-mode-expansion}. Equations~\eqref{eq:hor-common-weight}
and \eqref{eq:hor-radial-parity}, with $\eta=-1$ at $r_{\mathrm o}$ and $\eta=+1$
at $r_{\mathrm c}$, impose the physical future horizon conditions on
Eq.~\eqref{eq:ang-radial-system} for either sign of spin weight.

At the past black hole outer horizon, the outgoing
coordinates and the original tetrad are regular.
At the past cosmological horizon, the ingoing coordinates
and the boosted tetrad are regular.
Repeating the same argument therefore gives
Eqs.~\eqref{eq:hor-common-weight} and
\eqref{eq:hor-radial-parity} with $\eta=+1$ at
$r_{\mathrm o}$ and $\eta=-1$ at $r_{\mathrm c}$,
corresponding to the outgoing and ingoing conditions,
respectively.

The boundary identities also hold for complex $\omega$, without
conjugating the frequency; angular completeness has been established
only for real $\omega$. At resonant radial frequencies the Frobenius
recurrence may impose compatibility conditions or admit logarithmic
solutions. The full parity condition remains the regularity test,
but does not determine the existence or dimension of its solution
space at those frequencies.

\section{Summary and discussion}
\label{sec:discussion}
\label{sec:comparison}
\label{sec:kn-limit}
\label{sec:stu-comparison}

We have obtained an exact separation of the coupled gravitational
and electromagnetic radiative systems on a subextremal
Kerr--Newman--de Sitter background, under the assumptions stated
in Sec.~\ref{sec:geometry}.
The construction applies to both positive and negative spin weights
without expanding in the cosmological constant.
Its central ingredient is the explicit algebraic matrix
$\mathsf X$ in Eq.~\eqref{eq:explicit-X}.
It depends only on the background and is independent of
$\omega,m,\varsigma$.
Both $\mathsf X$ and its inverse extend regularly to the
horizons and angular poles in square-root boundary coordinates.
Building on Hintz's matrix integrating-factor
method~\cite{Hintz:2026jiw}, we use a transformation that
leaves the separated term
$(\alpha_r(r)+\alpha_\theta(\theta))\sigma_2$
in the transport equation.
This term can be incorporated into the radial and angular
equations separately, while the transformed coupling also
separates under Eq.~\eqref{eq:main-transformation}.
The resulting equations are the two-component angular system
\eqref{eq:ang-two-component} and radial system
\eqref{eq:ang-radial-system}, linked by the separation constant
$\lambda^{(\varsigma)}$.
Both systems are derived in Sec.~\ref{sec:target-transformation},
with $\alpha_r,\alpha_\theta$ given in
Eq.~\eqref{eq:global-connection-functions} and
$C_0^{(\varsigma)},\ldots,C_3^{(\varsigma)}$ in
Eqs.~\eqref{eq:ang-coefficient-combinations}--\eqref{eq:ang-coefficient-d}.

For real $\omega$ and integer $m$, the angular operator has a
complete orthonormal eigenbasis. Sec.~\ref{sec:separated}
shows that expansion in this basis reduces the transformed
radiative system to independent radial systems, one for each
angular eigenfunction.
The separated differential equations also hold for complex
$\omega$, although angular completeness is established here
only for real $\omega$.

Physical horizon regularity is preserved within each separated
pair. The future black hole outer and cosmological horizon conditions are
specified by the radial form \eqref{eq:hor-common-weight} and the
parity condition \eqref{eq:hor-radial-parity}, with $\eta=-1$
at $r_{\mathrm o}$ and $\eta=+1$ at $r_{\mathrm c}$, respectively.
These correspond to the ingoing condition at the black hole
outer horizon and the outgoing condition at the cosmological
horizon used to define quasinormal modes.
Sec.~\ref{sec:horizons} derives these conditions for both signs of spin
weight, including complex frequencies.
The inverse transformation in
Eq.~\eqref{eq:discussion-inverse-chiral} returns the radiative
variables; Eq.~\eqref{eq:discussion-Psi4} gives
$\Psi_4^{(1)}$ explicitly.

The assumptions $Q\ne0$, $a>0$, and distinct horizons are used,
respectively, by the phantom gauge and field normalizations, the
explicit matrix construction, and the square-root boundary coordinates.
We do not assert extensions through extremality or uniform
zero-charge and zero-rotation limits.
For fixed subextremal Kerr--Newman parameters,
Sec.~\ref{sec:local-kn-limit} relates the local
$\Lambda\to0$ limit to Hintz's formulation through
transformations that preserve separation.
The relation between regularity at the receding cosmological
horizon and the radiation condition at infinity remains
to be analyzed.

We also leave open compatibility of the systems for positive and
negative spin weights for one physical perturbation, and full
Einstein--Maxwell reconstruction.
The separated equations and horizon conditions provide
a starting point for quasinormal mode calculations and
stability analyses, which remain for future work.

We note that the local separation identities derived in
Sec.~\ref{sec:transformation} remain valid for
Kerr--Newman--anti-de Sitter after the replacement
$\ell^{-2}\mapsto-L^{-2}$, where $L>0$ and $0<a<L$.
For nonextremal black holes with $Q\ne0$, the same angular
and radial systems are therefore obtained on patches where
the transformation $\mathsf X$ is nonsingular.
Boundary conditions at anti-de Sitter infinity require
a separate analysis and are not considered here.

These results show that exact separation of the coupled
gravitational and electromagnetic radiative equations is not
peculiar to the asymptotically flat Kerr--Newman solution.
Its persistence in the presence of a positive cosmological
constant motivates a search for the geometric structure
underlying this separability.
The explicit form of $\mathsf X$ was found heuristically.
Developing a systematic procedure for constructing
such separation matrices directly from the radiative
equations on more general backgrounds is a further challenge.
It would be interesting to determine whether hidden
symmetries provide a geometric interpretation of the
transformation $\mathsf X$ and a route to generalizing it.
Natural settings include the Pleba\'nski--Demia\'nski
family~\cite{Plebanski:1976gy,Griffiths:2005qp,GriffithsPodolsky:2009}
and, at $\Lambda=0$, the
Ovcharenko--Podolsk\'y solutions with a nonaligned
electromagnetic field~\cite{Ovcharenko:2025cpm,Furugori:2026qbb}.
These backgrounds admit hidden
symmetries~\cite{Kubiznak:2007kh,Gray:2025lwy},
and Teukolsky-type gravitational equations have been separated
in the Dudley--Finley approximation~\cite{Dudley:1977zz,Dudley:1978vd,Gray:2025lwy}.
Extending this separation to the coupled gravitational
and electromagnetic system remains an open problem.
Whether Petrov type D alone is sufficient, or additional
conditions on the geometry and electromagnetic field are
required, remains open.
Identifying these conditions would clarify which backgrounds
admit separation of the coupled radiative system.

\section*{Acknowledgments}
This work was supported by JSPS KAKENHI Grant No.~JP26K07084.
The author thanks OpenAI for providing access to its AI tools through
the ChatGPT for Academic Researchers program.
The author used ChatGPT and Codex
(OpenAI; GPT-6 Astra)
to assist in searching for a suitable transformation matrix
through trial and error,
developing derivations, checking code for symbolic calculations,
and drafting and revising the manuscript.
The author made the scientific and editorial decisions
throughout the work and takes full responsibility
for the content of this work.

\appendix
\section{\texorpdfstring{Relation to Zhao's separation transformation}{Relation to Zhao's separation transformation}}
\label{app:zhao}

We collect the simpler separation matrix of
Ref.~\cite{Zhao:2026separation} and the associated matrices and
coefficient functions in the conventions of this paper.
These expressions can be used directly in
Eqs.~\eqref{eq:main-transformation}, \eqref{eq:ang-two-component},
\eqref{eq:ang-radial-system}, and \eqref{eq:ang-separated-ansatz}.
We then relate the two representations through the frame freedom
discussed in Sec.~\ref{sec:local-kn-limit}.
Throughout this appendix, we use the subextremal de Sitter
assumptions of Sec.~\ref{sec:geometry}.

\subsection{\texorpdfstring{The matrix and its auxiliary quantities}{The matrix and its auxiliary quantities}}
\label{app:zhao-matrix}

In our unit-determinant convention, Zhao's separation matrix is
\begin{equation}
 \mathsf X_{\mathrm Z}\coloneqq
 \frac{i}{\sqrt{\nu A_rD_\theta}}
 \begin{pmatrix}
 \dfrac{iMaA_r\cos\theta-(Mr-Q^2)D_\theta}{z}
 &\dfrac{iQ(vA_r-uD_\theta)}{z}\\[3mm]
 -\dfrac{iQ(vA_r+uD_\theta)}{w}
 &\dfrac{iMaA_r\cos\theta+(Mr-Q^2)D_\theta}{w}
 \end{pmatrix},\qquad
 \det\mathsf X_{\mathrm Z}=1.
 \label{eq:zhao-X}
\end{equation}
Here
\begin{equation}
 \eta_{\mathrm Z}\coloneqq
 \frac{2Q^2}{M^2-Q^2(1-\ell^{-2}a^2)+\nu},
 \label{eq:zhao-parameters}
\end{equation}
and
\begin{equation}
 A_r\coloneqq r^2-\eta_{\mathrm Z}a^2,\qquad
 D_\theta\coloneqq a^2(\cos^2\theta+\eta_{\mathrm Z}).
 \label{eq:zhao-denominators}
\end{equation}
The constant $\kappa$ of Ref.~\cite{Zhao:2026separation}, after
its de Sitter replacement, equals our $\nu$.
The matrix in Eq.~\eqref{eq:zhao-X}
is $iQ\sqrt\nu$ times the matrix in Eq.~(17) of
Ref.~\cite{Zhao:2026separation}, with the de Sitter replacements
of that reference.
This constant phase and normalization allow us to keep the
field transformation \eqref{eq:main-transformation} and the
coupling definition \eqref{eq:main-C-definition} unchanged in form.
When comparing mode coefficients, the frequency in that reference
is $\omega_{\mathrm{Zhao}}=\Xi\omega$, whereas $m$ is unchanged.
All expressions below use our $\omega$.

The corresponding auxiliary matrices are defined by
\begin{equation}
 \mathsf J_{\mathrm Z}\coloneqq
 \mathsf X_{\mathrm Z}\sigma_2\mathsf X_{\mathrm Z}^{-1}
 =\mathsf J,\qquad
 \mathsf W_{\mathrm Z}\coloneqq
 \mathsf X_{\mathrm Z}\mathsf X_{\mathrm Z}^\dagger,\qquad
 \mathsf E_{\mathrm Z}\coloneqq
 -\mathsf X_{\mathrm Z}\sigma_3\mathsf X_{\mathrm Z}^\dagger.
 \label{eq:zhao-auxiliary-definitions}
\end{equation}
In particular, $\mathsf J$ is exactly the matrix in
Eqs.~\eqref{eq:app-matrix-J}--\eqref{eq:app-matrix-J21}.
The other two matrices have the direct expressions
\begin{align}
 \mathsf W_{\mathrm Z}
 &=\mathsf I+\frac{2Qu}{\nu A_r}
 \begin{pmatrix}
 \dfrac{Q(uD_\theta-vA_r)}{\Sigma}
 &-\dfrac{MaA_r\cos\theta+i (Mr-Q^2)D_\theta}{z^2}\\[3mm]
 -\dfrac{MaA_r\cos\theta-i (Mr-Q^2)D_\theta}{w^2}
 &\dfrac{Q(uD_\theta+vA_r)}{\Sigma}
 \end{pmatrix},
 \label{eq:zhao-W}\\
 \mathsf E_{\mathrm Z}
 &=-\sigma_3+\frac{2Qv}{\nu D_\theta}
 \begin{pmatrix}
 \dfrac{Q(vA_r-uD_\theta)}{\Sigma}
 &\dfrac{MaA_r\cos\theta+i (Mr-Q^2)D_\theta}{z^2}\\[3mm]
 \dfrac{MaA_r\cos\theta-i (Mr-Q^2)D_\theta}{w^2}
 &-\dfrac{Q(vA_r+uD_\theta)}{\Sigma}
 \end{pmatrix}.
 \label{eq:zhao-E}
\end{align}
They satisfy the same algebraic identities as
$\mathsf J,\mathsf W,\mathsf E$ in
Eqs.~\eqref{eq:component-algebraic-identities}
and \eqref{eq:short-metric-identities}.

\subsection{\texorpdfstring{Separated coefficient functions}{Separated coefficient functions}}
\label{app:zhao-coefficients}

The transport equation takes the form
\begin{equation}
 \mathscr D\mathsf X_{\mathrm Z}
 =\mathsf A\mathsf X_{\mathrm Z}
 +\mathsf X_{\mathrm Z}
   (\alpha_{r,\mathrm Z}+\alpha_{\theta,\mathrm Z})\sigma_2,
 \qquad
 \alpha_{r,\mathrm Z}
 =\frac{M\eta_{\mathrm Z}a^2-Q^2r}{QA_r},\qquad
 \alpha_{\theta,\mathrm Z}
 =\frac{M\eta_{\mathrm Z}a^2}{QD_\theta}.
 \label{eq:zhao-alpha}
\end{equation}
These are the functions $h_r,h_\theta$ in Eq.~(20) of
Ref.~\cite{Zhao:2026separation}.
Keeping $K(r)$ and $W(\theta)$ as defined in
Eq.~\eqref{eq:bg-mode-potentials}, the coupling is
\begin{equation}
 \mathsf C_{\mathrm Z}^{(\varsigma)}
 \coloneqq-\mathsf X_{\mathrm Z}^{-1}
 \mathsf F^{(\varsigma)}(\mathsf X_{\mathrm Z}^\dagger)^{-1}
 =C_{0,\mathrm Z}^{(\varsigma)}\mathsf I
  +C_{2,\mathrm Z}^{(\varsigma)}\sigma_2
  +iC_{1,\mathrm Z}^{(\varsigma)}\sigma_1
  +iC_{3,\mathrm Z}^{(\varsigma)}\sigma_3,
 \label{eq:zhao-C}
\end{equation}
where
\begin{align}
 C_{0,\mathrm Z}^{(\varsigma)}
 &=\frac{3\Delta_r'}{4u}-\frac{3ru}{A_r}
   -\frac{i\varsigma K}{u}
   +\frac{2i\varsigma aQ^2u}{\nu A_r}\mathcal B_{\mathrm Z},
 \label{eq:zhao-C0}\\
 C_{2,\mathrm Z}^{(\varsigma)}
 &=-\frac{3\bigl(M(r^2+\eta_{\mathrm Z}a^2)-2Q^2r\bigr)}{2QA_r}
   +\frac{2i\varsigma aQ (Mr-Q^2)}{\nu A_r}\mathcal B_{\mathrm Z},
 \label{eq:zhao-C2}\\
 C_{1,\mathrm Z}^{(\varsigma)}
 &=\frac{3Ma^2(\cos^2\theta-\eta_{\mathrm Z})}{2QD_\theta}
   +\frac{2\varsigma a^2MQ\cos\theta}{\nu D_\theta}\mathcal B_{\mathrm Z},
 \label{eq:zhao-C1}\\
 C_{3,\mathrm Z}^{(\varsigma)}
 &=-\frac32(h\cot\theta+\partial_\theta h)
   -\frac{3av\cos\theta}{D_\theta}-\varsigma W
   -\frac{2\varsigma aQ^2v}{\nu D_\theta}\mathcal B_{\mathrm Z}.
 \label{eq:zhao-C3}
\end{align}
Here the common mode combination is
\begin{equation}
 \mathcal B_{\mathrm Z}\coloneqq
 \Xi\bigl(a(1+\eta_{\mathrm Z})\omega-m\bigr).
 \label{eq:zhao-mode-combination}
\end{equation}
These coefficients agree with Eq.~(18) of
Ref.~\cite{Zhao:2026separation} in our frequency convention:
$C_{0,\mathrm Z}=R_0$, $C_{2,\mathrm Z}=iR_1$,
$C_{1,\mathrm Z}=\Theta_1$, and $C_{3,\mathrm Z}=\Theta_0$.
They also follow directly by substituting
Eq.~\eqref{eq:zhao-X} into the transport and coupling definitions.
The calculation uses no complex conjugation of $\omega$ and
applies to both spin labels.
Replacing $\mathsf X$, $\alpha_r$, $\alpha_\theta$, and $C_j^{(\varsigma)}$
by these quantities in Sec.~\ref{sec:target-transformation}
therefore gives the separated systems for this representation.

\subsection{\texorpdfstring{Relation to the transformation in the main text}{Relation to the transformation in the main text}}
\label{app:zhao-relation}

The frame freedom discussed in Sec.~\ref{sec:local-kn-limit}
relates the two matrices by
\begin{equation}
 \begin{aligned}
 \mathsf X
 &=\mathsf X_{\mathrm Z}e^{-(f_{\mathrm Z}+ig_{\mathrm Z})\sigma_2}\\
 &=\mathsf X_{\mathrm Z}\bigl(
 \cosh(f_{\mathrm Z}+ig_{\mathrm Z})\,\mathsf I
 -\sinh(f_{\mathrm Z}+ig_{\mathrm Z})\,\sigma_2
 \bigr).
 \end{aligned}
 \label{eq:zhao-frame-relation}
\end{equation}
Here the real functions $f_{\mathrm Z}(r)$ and
$g_{\mathrm Z}(\theta)$ can be chosen as
\begin{align}
 e^{2f_{\mathrm Z}}
 &=\frac{\nu A_r(r^2+a^2)
 -2Qa^2(1+\eta_{\mathrm Z})u(Mr-Q^2-Qu)}
 {A_r\mathcal S_r},
 \label{eq:zhao-f}\\
 e^{2ig_{\mathrm Z}}
 &=\frac{Qa\cos\theta(k_\theta D_\theta-2v^2)
       +iMa^2v(\cos^2\theta-\eta_{\mathrm Z})}
       {D_\theta\sqrt{H_\theta}}.
 \label{eq:zhao-g}
\end{align}
The right-hand side of Eq.~\eqref{eq:zhao-g} has unit modulus.
Choose a continuous half-argument on $0\le\theta\le\pi$,
for example with $g_{\mathrm Z}(\pi/2)=-\pi/4$.
The otherwise arbitrary overall sign of $\mathsf X$ is fixed
so that Eq.~\eqref{eq:zhao-frame-relation} holds.
This transformation has the form of
Eq.~\eqref{eq:finite-lambda-integrating-factor},
and it preserves separation by the argument leading to
Eq.~\eqref{eq:finite-lambda-separated-coupling}.

One way to verify Eq.~\eqref{eq:zhao-frame-relation}
is to use the auxiliary matrices. Direct substitution gives
\begin{align}
 \mathsf W
 &=\cosh(2f_{\mathrm Z})\mathsf W_{\mathrm Z}
   -\sinh(2f_{\mathrm Z})\mathsf J\mathsf W_{\mathrm Z},
 \label{eq:zhao-W-relation}\\
 \mathsf E
 &=\cos(2g_{\mathrm Z})\mathsf E_{\mathrm Z}
   -i\sin(2g_{\mathrm Z})\mathsf J\mathsf E_{\mathrm Z}.
 \label{eq:zhao-E-relation}
\end{align}
Because both matrices $\mathsf X$ and $\mathsf X_{\mathrm Z}$
have determinant one and conjugate $\sigma_2$ to the same
$\mathsf J$, their ratio commutes with $\sigma_2$.
Equations~\eqref{eq:zhao-W-relation} and
\eqref{eq:zhao-E-relation} then fix its real and imaginary
exponents, respectively, leaving only a constant overall sign.

The change in the separated residual is
\begin{equation}
 \begin{aligned}
 \alpha_r
 &=\alpha_{r,\mathrm Z}-u\partial_r f_{\mathrm Z}
 -\frac{Qa^2(1+\eta_{\mathrm Z})}{M\ell^2},\\
 \alpha_\theta
 &=\alpha_{\theta,\mathrm Z}-h\partial_\theta g_{\mathrm Z}
 +\frac{Qa^2(1+\eta_{\mathrm Z})}{M\ell^2}.
 \end{aligned}
 \label{eq:zhao-residual-relation}
\end{equation}
The constant terms reflect the different division of the total
residual into radial and angular parts.
In contrast to the complete removal of the residual described
in Sec.~\ref{sec:local-kn-limit}, the inverse transformation
from $\mathsf X$ to $\mathsf X_{\mathrm Z}$ absorbs terms with
elementary primitives and leaves the simpler algebraic
functions in Eq.~\eqref{eq:zhao-alpha}.

%\clearpage

\bibliographystyle{JHEP}
\bibliography{ref}

\end{document}